\documentclass[sn-mathphys,Numbered,pdflatex]{sn-jnl}

\usepackage{graphicx}%
\usepackage{multirow}%
\usepackage{amsmath,amssymb,amsfonts,afterpage}%
\DeclareMathOperator*{\argmin}{arg\,min}
\usepackage{amsthm}%
\usepackage{mathrsfs}%
\usepackage[title]{appendix}%
\usepackage{xcolor}%
\usepackage{textcomp}%
\usepackage{manyfoot}%
\usepackage{booktabs}%
\usepackage{algorithm}%
\usepackage{algorithmicx}%
\usepackage{algpseudocode}%
\usepackage{listings}%
\usepackage{lineno}

\begin{document}

\title[Model Predictive Control of Neural Spiking]{Nonlinear Model Predictive Control of a Conductance-Based Neuron Model via Data-Driven Forecasting}


\author*[1]{\fnm{Christof} \sur{Fehrman}}\email{ckf5fh@virginia.edu}

\author[1,2]{\fnm{C. Daniel} \sur{Meliza}}\email{cdm8j@virginia.edu}

\affil*[1]{\orgdiv{Psychology Department}, \orgname{University of Virginia}, \city{Charlottesville}, \state{Virginia}, \country{United States of America}}

\affil[2]{\orgdiv{Neuroscience Graduate Program}, \orgname{University of Virginia}, \city{Charlottesville}, \state{Virginia}, \country{United States of America}}


\abstract{\textit{Objective}. Precise control of neural systems is essential to experimental investigations of how the brain controls behavior and holds the potential for therapeutic manipulations to correct aberrant network states. Model predictive control, which employs a dynamical model of the system to find optimal control inputs, has promise for dealing with the nonlinear dynamics, high levels of exogenous noise, and limited information about unmeasured states and parameters that are common in a wide range of neural systems. However, the challenge still remains of selecting the right model, constraining its parameters, and synchronizing to the neural system. \textit{Approach}. As a proof of principle, we used recent advances in data-driven forecasting to construct a nonlinear machine-learning model of a Hodgkin-Huxley type neuron when only the membrane voltage is observable and there are an unknown number of intrinsic currents. \textit{Main Results}. We show that this approach is able to learn the dynamics of different neuron types and can be used with MPC to force the neuron to engage in arbitrary, researcher-defined spiking behaviors. \textit{Significance.} To the best of our knowledge, this is the first application of nonlinear MPC of a conductance-based model where there is only realistically limited information about unobservable states and parameters.}


\maketitle
\section{Introduction}\label{sec1}
\subsection{Control of Neural Systems}
Precise control of neural systems is a major goal of modern neuroscience, both as a means for experimental investigation of the brain and as a clinical method for treating neurological disorders \cite{schiff2011neural}. In the most general sense, we seek to find a command signal that will force a specific neuron or network of neurons to follow a specified trajectory through state space \cite{kao2019neuroscience}. More informally stated, how can we make the system do what we want it to do? This level of control would allow us to experimentally test the predictions of hypotheses in neural systems and potentially restore normal function to a circuit that has gone into a pathological state \cite{parkes2023using}.

\begin{figure}
    \centering
    \includegraphics[width=0.9\textwidth]{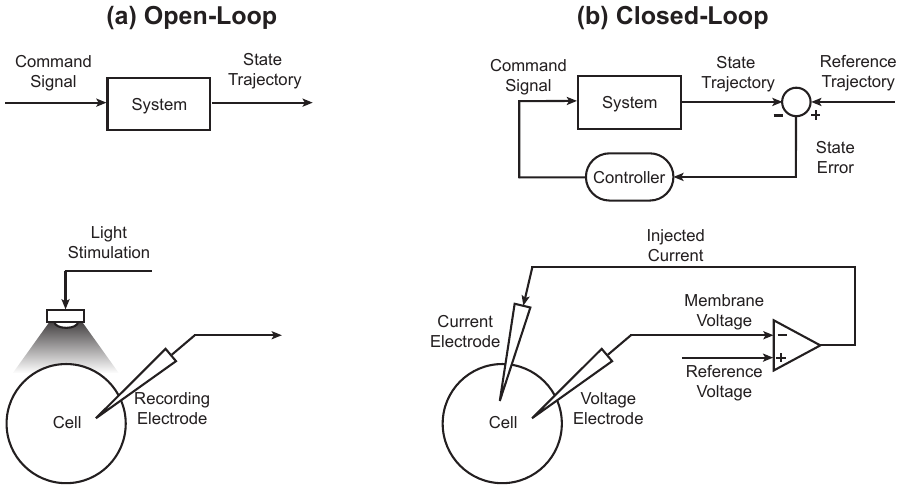}
    \caption{\textbf{Open- vs Closed-Loop Control}.
    (a) (Above) Block diagram of open-loop control. A command signal is applied to the system irrespective of the system output. (Below) Diagram of typical optogenetic stimulation experiment. A light source is applied to a cell expressing the appropriate light-sensitive opsin. For open-loop stimulation, the intensity of the light is determined before recording and may or may not cause the cell to fire. (b) (Above) Block diagram of closed-loop control. A command signal is calculated by the controller based on the state error - the difference between the system state trajectory and reference trajectory. (Below) Diagram of a voltage-clamp experiment. The cell is held at a specified voltage by injecting current determined by an online comparison of the membrane voltage with the desired reference voltage.}
    \label{fig:Figure 1}
\end{figure}

Achieving control presupposes the ability to predictably manipulate the system. In open-loop control, a command signal is chosen ahead of time based on a general model of the system and then applied to an individual instance (Figure~\ref{fig:Figure 1}(a)). The outcome may inform the general model, but this occurs offline. In neurophysiology, examples of open-loop control include current-clamp intracellular recording as well as most optogenetics experiments, where a pulse of current or light is used as the command signal to force a neuron to spike or prevent it from spiking \cite{emiliani_optogenetics_2022}. What makes these open-loop is that the intensity and duration of the pulse is not automatically adjusted if the stimulus fails to achieve the desired effect \cite{grosenick_closed-loop_2015}. Although open-loop control has the advantages of being fast and simple to implement, it is not robust to unknown disturbances or errors in command signal calculation \cite{zaaimi_closed-loop_2022}. Because neurons \emph{in vivo} receive many spontaneously active excitatory and inhibitory inputs, there may be significant trial-to-trial variability in the number of spikes evoked during application of the command signal. More broadly, variability in the actual effects of a manipulation reduces the power to make causal inferences in experimental settings.

In contrast, for closed-loop (or feedback) control, the command signal is dynamically adjusted as a function of the difference between the actual (or estimated) state of a specific system and the desired reference trajectory (Figure~\ref{fig:Figure 1}(b)). This is the approach employed in voltage-clamp experiments, where the difference between the actual and the desired membrane voltage (the state error) is scaled by a gain factor and used as the command signal of electrical current injected into the neuron \cite{jaeger_voltage-clamp_2014}. Closed-loop controllers have the ability to adapt to unknown system disturbances and changes in system dynamics even when tracking complicated reference trajectories \cite{stefani_design_2002}. Because of this, considerable work has gone into incorporating feedback controllers in other areas of neuroscience such as brain-machine interfaces \cite{shanechi_rapid_2017, gilja_brain_2012, willett_feedback_2017, zhang_prototype_2021} and neuro-prosthetics \cite{shanechi_robust_2016,shanechi_rapid_2017,cunningham_closed-loop_2011,wright_review_2016,pandarinath_science_2022,pedrocchi_error_2006}. In particular, there have been recent advancements in using feedback controllers with optogenetic stimulation to give more fine-tuned and reliable control of neural spiking at both the individual neuron \cite{bolus_state-space_2021} and population levels \cite{bergs_all-optical_2023,newman_optogenetic_2015}.

Although a promising avenue of research, there are still many issues when using feedback controllers with complicated systems. Most implementations of feedback control are purely \textit{reactive}, where the command signal is a function of the present and/or past state error terms. Reactive control works exceptionally well with intracellular preparations that allow low-noise measurements of voltage and direct injection of current \cite{jaeger_voltage-clamp_2014}; indeed, voltage-clamp experiments are foundational to almost all of what we know about the physiology of individual neurons. However, there are significant obstacles to using reactive control in networks of neurons, which can have highly nonlinear dynamics in much larger state spaces, more unobservable states and parameters, and proportionally fewer variables that can be experimentally controlled. Our goal in this study is to explore the application of \textit{anticipatory} control to neural systems, using a single neuron with Hodgkin-Huxley dynamics as a proof of principle. Although this is a simple system that does not need sophisticated methods to control it, we are able to use it to address one of the major problems likely to arise when applying more sophisticated methods to complex circuits, namely the lack of ground-truth knowledge about the dynamics and the unmeasureable states of the system.

\subsection{Model Predictive Control}
One promising method of feedback control to deal with these problems is model predictive control (MPC), which is a type of optimal controller. It is optimal in the sense that the control input $u$ minimizes an objective function of the form

\begin{equation}
    J(x_0) = \sum_{i=0}^{T} \ell (x_i,u_i), \\
\end{equation}

\noindent
with constraints

\begin{align}
    x_{n+1} & = f(x_n,u_n)\nonumber \\
    x_{LB} & \le x \le x_{UB}\nonumber \\
    u_{LB} & \le u \le u_{UB}\nonumber,
\end{align}

where $\ell (x_i,u_i)$ is the loss associated with $i$th time step, which is a function of the state variable(s) $x$ and input(s) $u$. Many types of loss functions are possible, but typically involve the state error and energy cost of the command signal. The constraints allow one to specify the dynamics of the system and to give lower and upper bounds for the state variables and inputs. More sophisticated versions of MPC allow for additional constraints where knowledge of any measurement or process noise can be incorporated \cite{hewing_learning-based_2020}, but this will not be explored here. 

\begin{figure}
    \centering
    \includegraphics[width=0.9\textwidth]{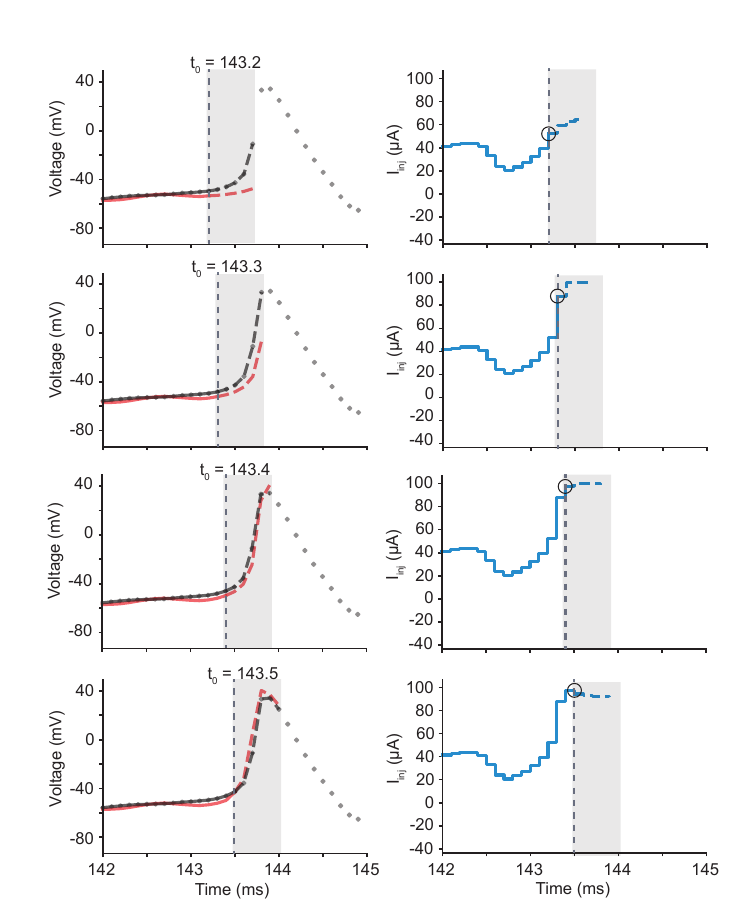}
    \caption{\textbf{Receding Horizon of MPC.} Starting at the top row, (left) the system state trajectory (red) is being controlled to follow the reference trajectory (black). At the current time step (vertical dotted line) the controller finds the optimal set of inputs that minimize the loss function for a specified time horizon (shown with a gray box). In this case, the controller looks ahead 5 time steps (black dashed curve). Given the state at the current time ($t_0$), the controller uses a model to predict where the system will be across the future time horizon (red dashed curve). The inputs into the system (right) are optimized in discrete-time, and the input into the system is held constant between model time steps (solid blue curve). The predicted optimal future inputs (dashed blue curve) are calculated across the future time horizon. However, only the first of these values (circled in black) is used as input in the next time step before the optimization procedure begins again. From the top to bottom rows, we see how the controller may pick new optimal inputs given updates in the model predictions and by having access to new reference trajectory values (black dotted curve).}
    \label{fig:Figure 2}
\end{figure}

The controller uses a discrete-time model of the system $f(x_n, u_n)$ to predict what command inputs would best force the system to follow the reference trajectory over some time horizon $T$ (Figure~\ref{fig:Figure 2}). At each time step, the controller finds an optimal command signal by minimizing the total loss given the constraints. The total loss is calculated by summing the actual and predicted losses across the time horizon. However, only the first time step in the optimized control signal is applied to the system, and the optimization is performed again in the next time step. This process repeats at each discrete time step, which leads some to refer to MPC as receding horizon control \cite{holkar2010overview}. By finding an optimal input based on predictions of how the system will behave in the future, MPC is an \textit{anticipatory} controller \cite{lin_development_2023}. Although the command signal is only guaranteed to be globally optimal for linear systems with convex loss functions, MPC has been widely used in nonlinear system control \cite{rakovic_handbook_2019,brunton_data-driven_2019}.

A commonly used analogy to describe MPC is the game of chess \cite{rakovic_handbook_2019}, where the player (the controller) wants to find a set of moves (the command signal) to win the game (the objective function). When selecting a move, the player must use a model of their opponent to anticipate how that opponent will respond to their moves. Although the player may have mapped out their moves for the next $T$ turns (the time horizon), the player can only implement the first of these moves during their turn. The player may update their planned moves based on a variety of factors. Their opponent may have selected a different move than predicted, or after completing their turn the player is able to think one more move ahead (the receding horizon) and finds a new optimal set of moves. Intuitively, being able to think more moves ahead (extending the length of the time horizon) should produce a more optimal set of moves to win the game. However, this comes at the cost of increased computational complexity for the player, and errors in modeling how the opponent will respond can accumulate when incorporating these errors across the extended time horizon. This leads to a balancing act in MPC where not having a large enough time horizon may result in suboptimal moves in the long run, whereas too long of a time horizon is expensive and sensitive to modeling errors.

One of the primary considerations in implementing MPC is choosing a good model of the system one wants to control \cite{schwenzer_review_2021}. At first glance, this might not seem to be a problem for applications in neuroscience since constructing mathematical models of neural systems is one of the main research areas. For example, models based on voltage-dependent ionic conductances using the Hodgkin-Huxley framework can accurately predict how the membrane voltage of a neuron with a given morphology and complement of currents will respond to an arbitrary input. However, building a conductance model of a specific neuron is far from trivial \cite{rabinovich_dynamical_2006}. The types of currents must be chosen along with dozens to hundreds of free parameters that govern the maximal conductances of the intrinsic currents and their voltage-dependent kinetics, and there are many state variables of which only the membrane voltage is typically observable \cite{toth_dynamical_2011}. MPC has been successfully applied to conductance models in previous work \cite{frohlich_feedback_2005,yue_non-linear_2022,senthilvelmurugan_active_2023}, but always with the assumption that the number, type, and/or functional forms of all the intrinsic currents are known \textit{a priori}. Under this assumption, there are many data assimilation methods that can be used to estimate the unknown parameters and hidden states of the model \cite{toth_dynamical_2011,kostuk_dynamical_2012,ullah_tracking_2009}. However, in most biological preparations, this is an unrealistic assumption, because neurons express a large, diverse complement of voltage-gated channels whose physiological properties can depend on variations in isoform composition, modulatory subunits, phosphorylation state, and subcellular localization. Choosing even a general form of a model to be used with a specific type of cell requires significant hand-tuning \cite{Meliza:2014p15146} that would not be feasible if the goal is to control a specific neuron or network in a live experiment.

\subsection{Data-Driven Approaches to Modeling}
An alternative to constructing a detailed biophysical model is to use a data-driven approach where the dynamics of the system are modeled based on empirical data with minimal reference to the underlying biology of the neuron. Using standard machine-learning approaches, unknown parts of the system can be modeled via function approximation and used to predict the time-evolution of the system to various inputs. These models are often referred to as forecasting models since the model predicts how the system will change across time. Data-driven approaches have been successfully applied to MPC problems in diverse fields \cite{bieker_deep_2019,kaiser_sparse_2018,hewing_learning-based_2020,salzmann_real-time_2023,zheng_physics-informed_2023}. While there has been previous work using these approaches to model HH-type neuron models, the models were either used solely for prediction (instead of control) \cite{plaster_data-driven_2019} or made unrealistic assumptions about which state variables were available in the training data or the extent to which the complement and functional forms of the intrinsic currents could be known \cite{yue_non-linear_2022,senthilvelmurugan_active_2023}.

In order for data-driven models to be useful for MPC applications in neuroscience, these models must be able to accurately predict the states to be controlled based only on observable state measurements, be agnostic to the number of hidden states and intrinsic currents, and generalize to a control scheme where command signals may be outside the training set. As a proof of principle, we conducted a simulation study to control the membrane voltage of an HH-type neuron through current injection when the parameters of the model were unknown, and only the membrane voltage was observable. We used these observations to create a nonlinear data-driven model that accurately predicted the response of the system to command signal inputs and used this model for MPC. The model made no assumptions about the nature of the intrinsic currents and still allowed the controller to force the membrane voltage to follow a reference trajectory. Although control of single unit voltage activity is achievable with proportional feedback control both \emph{in vivo} and \emph{in vitro} \cite{sherman2012axon}, our goal was to demonstrate how data-driven modeling can be applied to nonlinear MPC of neural systems. To our knowledge, this is the first application of nonlinear MPC to a spiking neuron model where realistically limited knowledge of the system is known.

\section{Methods}
\subsection{Connor-Stevens Model}
As a model of single-unit responses to an injected current, we used the HH-type Connor-Stevens (CS) model for all simulations \cite{sterratt_principles_2011}. The CS model includes four intrinsic currents and an extrinsic injected current. We also included a noise current that modeled the unknown, variable synaptic inputs that contribute to the trial-to-trial variability neurons tend to exhibit \emph{in vivo}. The model is given by the equations
\begin{equation}
    C\frac{dV}{dt} = I_{Na}+I_{K}+I_{A}+I_{l}+I_{noise}+I_{inj},
\end{equation}
where
\begin{align}
\label{connor-stevens}
        I_{Na} & = g_{Na} m^3 h (E_{Na}-V)\nonumber \\ 
        I_K & = g_K n^4 (E_K-V)\nonumber \\
        I_A & = g_A a^3 b (E_A-V)\nonumber \\
        I_l & = g_l(E_l-V)\nonumber,
\end{align}
where $I_{Na}$, $I_K$, and $I_{A}$ are the voltage-dependent sodium, potassium, and A-type intrinsic ionic currents, and $I_l$ is the intrinsic leak current. The activity of the neuron can be externally modulated by varying the injected current $I_{inj}$. For the noise current $I_{noise}$, random Poisson spike trains were convolved with an alpha function with a decay rate of $10$ ms \cite{sterratt_principles_2011} to model the resulting post-synaptic potentials. Each CS model received inputs from both an excitatory and inhibitory Poisson neuron with a firing rate of $20$ Hz. The amplitude of $I_{noise}$(t) was scaled in reference to the training and validation injected currents to have a constant signal-to-noise ratio (SNR) of 5. Note that all currents are functions of time but we omit making this explicit in the equations for simplicity. Each of the three voltage-gated currents depend on one or more unobservable state variables that model the activation state ($m,n,a$) and inactivation state ($h,b$) of the channels. Each of these state variables is governed by a first-order differential equation with unique parameters that determine its kinetics. While in principle one could estimate the values of the state variables and model parameters using data assimilation techniques \cite{toth_dynamical_2011,kostuk_dynamical_2012,knowlton_dynamical_2014}, in an actual biological preparation one would be unlikely to know all of the channels a specific neuron expresses.

\begin{figure}
    \centering
    \includegraphics[width=1\textwidth]{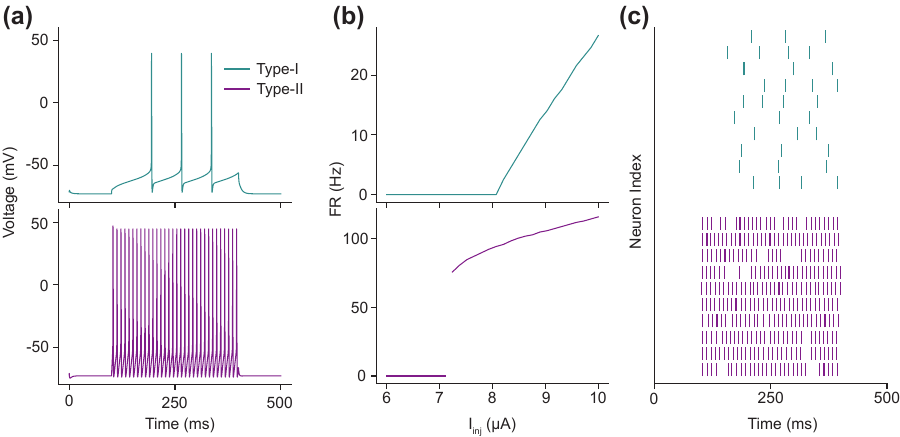}
    \caption{\textbf{CS Model Behaviors.} (a) The spiking pattern of a Type-I (above) and Type-II (below) CS model in response to a 300 ms 9 $\mu$A step current. (b) The firing rate of the CS model as a function of step current amplitude. Notice that the Type-I model's firing rate increases approximately linearly after the input passes the firing threshold while the Type-II model abruptly jumps in firing rate. (c) The effect of the noise current on the CS models when stimulated with the same step current from panel A.}
    \label{fig: Figure 3}
\end{figure}

The CS model is able to produce distinct firing dynamics by changing the parameter $E_l$ and $g_A$ parameter values (Figure~\ref{fig: Figure 3}). With $g_A = 47.7 \text{ mS}$ and $E_l = -22$ mV, the model exhibits Type-I excitability, which is characterized by a smooth increase in firing rate when the input currents exceed the firing threshold. When $g_A = 0\text{ mS}$ (eliminating the A-type current) and $E_l = -72.8$ mV, the model instead exhibits Type-II excitability, which is characterized by a discontinuous jump in firing rate for input currents that exceed the firing threshold. This difference in spiking behavior reflects two distinct dynamical topologies that undergo qualitatively different kinds of bifurcations: Type-I spiking is indicative of a saddle-node bifurcation, whereas Type-II spiking is caused by an Andronov-Hopf bifurcation. To show that data-driven approaches to MPC can extend to various types of neural dynamics and number of intrinsic currents, we used both the Type-I and Type-II CS models in all simulated experiments.

\subsection{Data-Driven Forecasting of CS Model}
The HH model and its variants are conductance-based models where the cell membrane is modeled as a capacitor \cite{skinner2006conductance}. Thus, the relationship between the membrane voltage and cellular currents can be expressed using the current conservation equation
\begin{equation*}
    C\frac{dV}{dt} = \sum_i I_i(t),
\end{equation*}
where the time derivative of the membrane voltage $V$ is proportional to the sum of all currents through the membrane. These currents may be externally applied (e.g. electrode injected currents) or intrinsic to the neural dynamics themselves (e.g. arising from voltage- and ligand-gated ion channels). To construct a forecasting model, we only assumed the dynamics of the membrane voltage were given by
\begin{equation}
    C\frac{dV}{dt} = F(V,X,\Theta,t)+I_{inj},
    \label{eq:ddf-dvdt}
\end{equation}
where $C$ is the membrane capacitance and $F(.)$ is an unknown time-varying function of membrane voltage, with unknown states $X$ and unknown parameters $\Theta$. We stress that this assumption would hold not only for the CS model, but any conductance-based model because of the additivity of the currents. For the CS model, this $F(.)$ would be the intrinsic currents, $X$ would be the intrinsic state variables, and $\Theta$ would be the model parameters. The goal in data-driven forecasting (DDF) is not to estimate these unknowns but to approximate $F$ such that one can accurately predict how the neuron will respond to an arbitrary input current $I_{inj}$ by integrating Equation (\ref{eq:ddf-dvdt}).

Let $\mathbf{V} = [V_0, V_1, ..., V_T]$ denote a set of discretely sampled membrane voltages where $V_k = V(k\Delta t)$ and $\Delta t$ is the sampling period. Similarly, let $\mathbf{I} = [I_0,I_1,...,I_T]$ be the set of discretely sampled injected currents. Given only $\mathbf{V}$ and $\mathbf{I}$, the goal is to find a DDF model of the form $V_{n+1} = F_{DDF}(\mathbf{V},\mathbf{I})$ that can accurately map $V_n$ to $V_{n+1}$. There are many possible models $F_{DDF}(.)$ to choose from and as a general rule demand larger amounts of training data as the DDF model gets more complex \cite{bourdeau_modeling_2019}. Additionally, if one used both the membrane voltage and injected currents as input into black-box function approximator, it would be difficult to separate the effects of the intrinsic dynamics of the system (membrane voltage) from the effects of the external force (injected current). An approach taken by \cite{clark_reduced-dimension_2022} when modeling the dynamics of HH-type neurons was to exploit the fact that the $I_{inj}$ term is additive and remove that from the function approximation step. In \cite{clark_reduced-dimension_2022}, they were able to achieve a good forecasting model by using time-embeddings of $\mathbf{V}$ in conjunction with a radial basis function network (RBFN) \cite{lowe1988multivariable}. This is a single hidden-layer artificial neural network (ANN) that typically uses a Gaussian as the nonlinear activation function. Although this is a classical approach largely superseded by more modern forecasting models such as LSTMs \cite{hochreiter1997long} and Transformers \cite{vaswani2017attention}, their DDF model generalized to \textit{in vitro} recordings across many different neuron types. In contrast to more complex models, RBFNs are easier to train while still being universal function approximators \cite{park_universal_1991}. The general form of this DDF model is given by the equation
\begin{equation}
    V_{n+1}= V_n + F_{RBF}(S_n)+\alpha(I_{n+1}+I_n),
\end{equation}
where $V_n$ is the membrane voltage at the $n$th time sample, $S_n$ is a time-embedding of $V_n$, $F_{RBF}(.)$ is a RBFN with learned parameters, $I_n$ is the injected current at the $n$th time sample, and $\alpha$ is a learned scaling parameter. See Appendix for a brief derivation of the DDF model and \cite{clark_reduced-dimension_2022,clark_data_2022} for a more detailed treatment.

\subsubsection{Simulating Training Data}
Separate DDF models were trained on data from the Type-I and Type-II CS neuron models. The injected currents used to stimulate the CS neurons were obtained from the $x(t)$ state of the Lorenz 63 system \cite{lorenz1963deterministic}. This chaotic current has been shown to cover a large frequency spectrum and has been used to drive \emph{in vitro} neurons across a sufficient extent of their state space to support accurate data assimilation \cite{toth_dynamical_2011}. The differential equations governing the system were scaled in time by ($\mathbf{\tau}$) and the resulting $x(t) $trajectory in amplitude ($\mathbf{A}$):
\begin{align}
\mathbf{\tau}\dot{x} &= \sigma(y - x) \\
\mathbf{\tau}\dot{y} &= x(\rho - z) - y \\
\mathbf{\tau}\dot{z} &= xy - \beta z
\end{align}
\begin{equation}
    I_{inj}(t) = \mathbf{A}x(t)
 \end{equation}
 where $\mathbf{\tau} = 20$ and the amplitudes for the CS model Type-I and Type-II models were $1.8$ and $0.5$ respectively. 

All simulations were performed using \verb|scipy.integrate.odeint| with a time window $h = 0.02$ ms. Five seconds of simulated data were used as training data for each of the DDF models. Because MPC is computationally expensive, we would not expect to be able to run the optimization process (Figure \ref{fig:Figure 5}, blue loop) at the sampling rate of the recording, and so we down-sampled the membrane voltage and injected currents to 10 kHz, which corresponds to a sampling period of $\Delta t = 0.1$ ms. This is a relatively low sampling rate for voltage-clamp experiments and demonstrates we are still able to control these systems with less data than typically used to build biophysical conductance-based models.

\subsubsection{RBFN Hyperparameters}
A Gaussian was used as the radial basis function in the RBFNs, which is of the form
\begin{equation}
    \psi_c(S_n) = \exp\{-R||S_n-\mu_c||^2\}.
\end{equation}
The dimension of the time-embedding was chosen using the Simplex method as described in \cite{sugihara1990nonlinear} with the \verb|pyEDM| package. Using a time delay $\tau^*$ of 1, the optimal predictive embedding dimension $D_e$ was found to be $2$ for both DDF models (i.e., $S_n = [V_n,V_{n-1}]$). The center vectors $\mu_c$ (N = 50) were obtained by performing k-means clustering in the time-embedded space of the training data. For all RBFs, a length scaling parameter of $R = 0.01$ was used.

\subsubsection{RBFN Training}
The RBFNs were trained via Ridge regression (also known as Tikhonov regularization) with the solution given by
\begin{equation}
    W = (X^T X+\lambda I)^{-1}X^T Y,
\end{equation}

where
\begin{equation}
     Y = \begin{bmatrix}
           V_1-V_0 \\
           V_2-V_1 \\
           \vdots \\
           V_T-V_{T-1}
         \end{bmatrix}, X =
          \begin{bmatrix}
           \psi_1(S_0) &  \psi_2(S_0) & \hdots & \psi_N(S_0) & I_1+I_0\\  
           \psi_1(S_1) &  \psi_2(S_1) & \hdots & \psi_N(S_1) & I_2+I_1\\  
           \vdots & \vdots & \ddots & \vdots & \vdots \\
           \psi_1(S_{T-1}) &  \psi_2(S_{T-1}) & \hdots & \psi_N(S_{T-1}) &I_T+I_{T-1}\\
          \end{bmatrix}, W =
          \begin{bmatrix}
           w_{1} \\
           w_{2} \\
           \vdots \\
           \alpha
         \end{bmatrix}.
\end{equation}
Model training was performed using the $sklearn$ python package \cite{pedregosa2011scikit} with 10-fold cross-validation to obtain an optimal $\lambda$ regularization parameter.

\subsubsection{DDF Model Evaluation}

\begin{figure}
    \centering
    \includegraphics[width=1\textwidth]{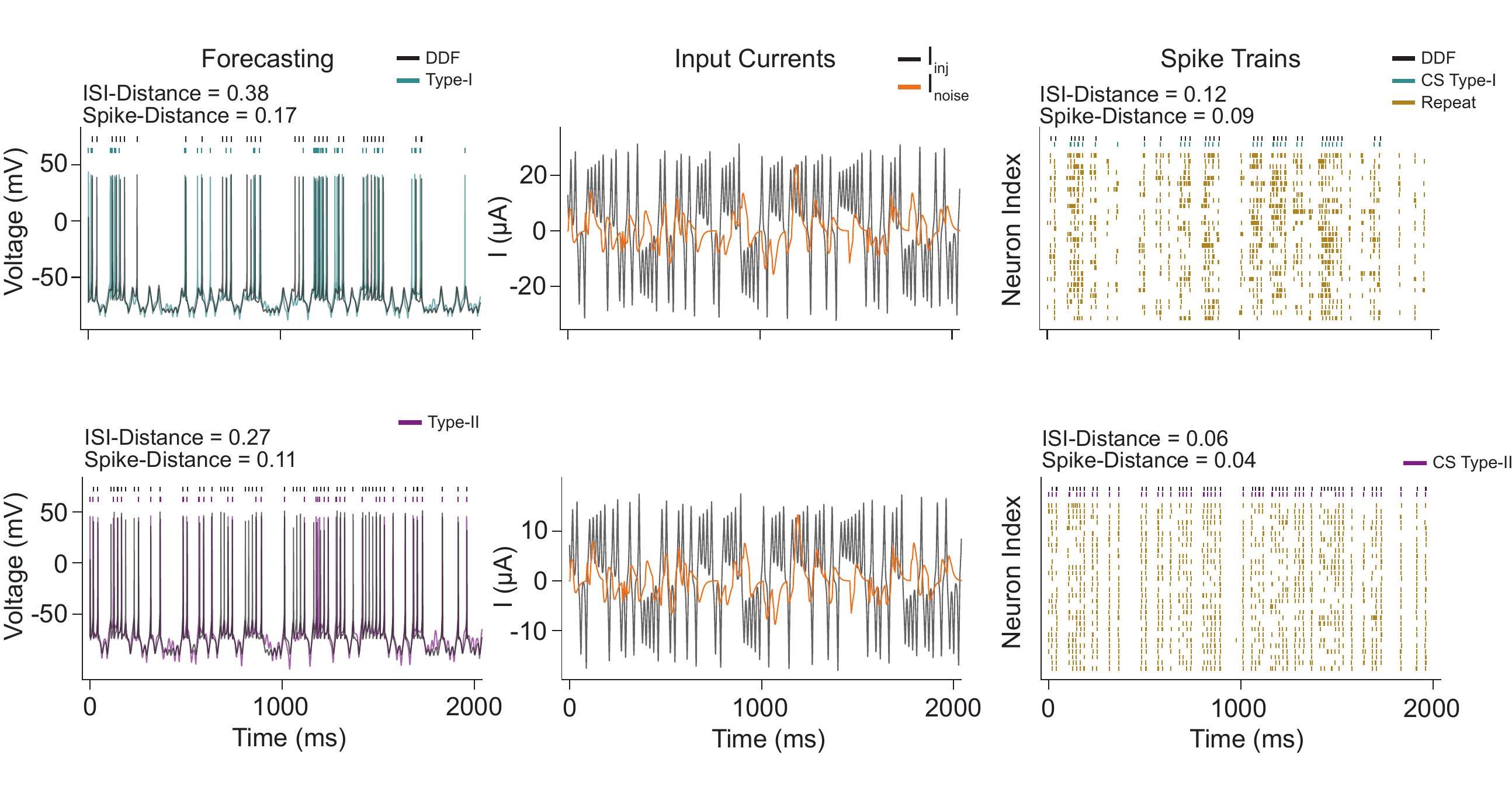}
    \caption{\textbf{DDF Model Forecasts.} Each CS model was stimulated with 5 seconds of a known $I_{inj}(t)$ and unknown $I_{noise}(t)$. Each of the DDF models were fit using only the observations of the CS model voltages and the $I_{inj}(t)$. To evaluate the fit of the DDF models, their forecasted membrane voltages and spike trains were compared to CS models stimulated by a validation set of injected and noise currents. The forecasts were completely open-loop, where the DDF model was not corrected based on errors in predictions. (Left) CS model state trajectories (colored) and DDF model forecasts (black) on 2000 ms of validation data. (Middle) The $I_{inj}(t)$ (black) and $I_{noise}(t)$ (orange) currents for the section of validation data. The injected currents used to train and validate the model were obtained using the chaotic Lorenz 63 system. Poisson neurons were used to produce the noise current and had balanced excitation and inhibition. The amplitudes of the noise currents were chosen to result in an SNR of 5 compared to the injected current. This can be seen in the differing scales of the injected and noise currents between the two models. (Right) In black are the DDF forecasted rasters and directly below are the CS model spikes when only stimulated by $I_{inj}(t)$. We see a strong similarity between the two spike trains indicating that the DDF model learned much of the CS model dynamics. Although the DDF model did not accurately forecast every spike in the validation data (a), this was largely due to the unknown sources of noise. Repeated simulations of the CS model with the same injected current but different noise currents (yellow) show that the CS model without $I_{noise}(t)$ is deficient in predicting the noisy spike trains.}
    \label{fig:Figure 4}
\end{figure}

\begin{figure}
    \centering
    \includegraphics[width=0.9\textwidth]{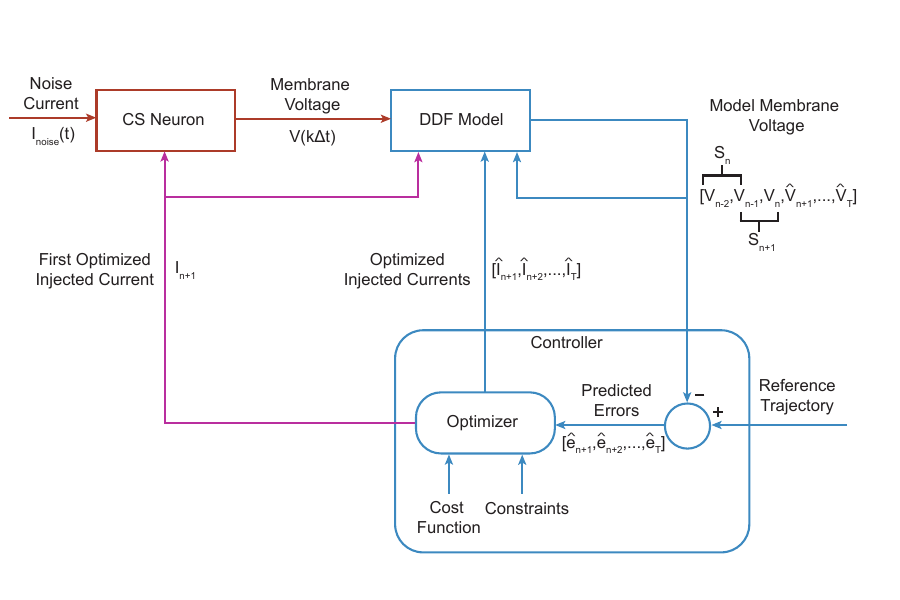}
    \caption{\textbf{MPC via DDF Diagram.} (Red) The CS neuron receives an input $I_{inj}(t)$ from the controller at time step \textit{n} which is held constant across a time interval of $\Delta t$ seconds. (Blue) The DDF model gets an update of the CS model membrane voltage every $\Delta t$ seconds. Given the membrane voltage $V_n$, time-embedded state history $S_n$, and discrete-time input $I_n$, the controller finds an optimal input for the next time step $I_{n+1}$. Given this optimized input, the DDF model makes a prediction of the CS model membrane voltage at the next $\Delta t$ time step, $V_{n+1}$. The controller uses the DDF model to simulate 5 $\Delta t$ time steps into the future (the time horizon) to find a sequence of optimized inputs by minimizing the loss function. (Purple) The first of these optimized inputs $I_{n+1}$ is used as the next injected current into the CS neuron. }     
    \label{fig:Figure 5}
\end{figure}

Although previous work has shown that the DDF model has high accuracy for \textit{in silico} and \textit{in vitro} neurons \cite{clark_reduced-dimension_2022}, the sampling rate was much higher than our data ($\geq 50$ kHz). To assess whether DDF would work on the CS model using data with a lower sampling rate more in line with the control loop speed we might expect to achieve in a live, biological preparation, we performed open-loop forecasting on novel injected currents. Because the DDF models had a time-embedding parameters $D_e = 2$ and $\tau^* = 1$, the first two $\Delta t$ time samples were used to seed the model. As seen in Figure~\ref{fig:Figure 4}, the DDF model was able to accurately predict the response of the CS neuron to a novel current injection. Although the DDF model did not accurately forecast every spike, it is important to note that it was only able to use the injected current to make predictions of the time-evolution of the system. We believed this was largely due to the unknown synaptic noise current rather than significant errors in the model. To test this, we compared the DDF predictions to the responses of the original CS neuron to noisy currents (Figure~\ref{fig:Figure 4}, right). By running the CS model with different instantiations of the noise current, we can see that the prediction error of the DDF model is comparable to the variability due to the noise current.

\newpage
\subsection{Model Predictive Control using a DDF Model}
Controlling the membrane voltage of a CS neuron via MPC was performed by finding an optimal set of injected current inputs that minimized the cost function
\begin{equation}
    \argmin_{I_1,I_2,...,I_T} \mathbf{s}e_T^2+ \sum_{n=0}^{T-1} \mathbf{q}e_{n+1}^2 + \mathbf{r} \Delta I_{n+1}^2,
    \label{eq:mpc_loss}
\end{equation}
subject to the constraints
\begin{equation*}
    V_{n+1}= V_n + F_{RBF}(S_n)+\alpha(I_{n+1}+I_n)
\end{equation*}
\begin{equation}
    |I_n| \leq 100 \: \mu A,
\end{equation}
    
where $V_0$ is the membrane voltage at the current time step, $e_n$ is the error between the membrane voltage $V_n$ and the reference trajectory $V^{ref}_n$ at the $n$th time step relative to $V_0$, and $\Delta I_{n+1} = I_{n+1}-I_{n}$ (also relative to $V_0$). Note that the $\argmin$ corresponds to controlling the $I_{n+1}$ term in the DDF model. For the very first optimization loop, we set $I_0$ to 0 $\mu$A. The controller hyperparameters $\mathbf{s},\mathbf{q}$ and $\mathbf{r}$ allow one to differentially weight errors in control and errors in input fluctuations. Setting $\mathbf{r} = 0$ can result in rapid input fluctuations which may make the controller perform poorly \cite{qin_survey_2003}. One could additionally add another term to the cost function that penalizes the squared magnitude of the input current.

At the beginning of the control loop, the controller uses a model of the system to simulate $T$ time steps into the future in order to find the optimal set of inputs to minimize the cost function. Recall that the DDF model is working in 0.1 ms time steps resulting in the control input $I_{inj}$ applied to the CS neurons being kept constant for that time window. Reducing the width of this window would enable one to control systems at faster time scales but at the cost of increased computational load.

While in principle, one could measure the controlled system once to get the initial values and use the data-driven model forecasts as an estimate of the actual state for the entirety of the control loop, we update the DDF model at every sampling period with the corresponding membrane voltage values of the CS neuron. This was done due to the high amount of noise present in the system. In systems were there is a low amount of noise, data-driven models that accurately forecast the dynamics could be used without the need of the constant monitoring of the system states.

All MPC optimizations and implementations were performed using the \verb|do-mpc| python package \cite{fiedler_-mpc_2023}. This package utilizes \verb|CasADi| \cite{andersson_casadi_2019} and \verb|IPOPT| \cite{wachter_implementation_2006} for optimization and automatic differentiation methods. See Appendix for controller hyperpameters.

Note that the range in which the control input operates is higher than is typical of patch-clamp experiments \cite{sherman2012axon}. The CS model has parameters that are normalized by soma surface area which result in different conductance and capacitance values when modeling cells of different sizes. For simplicity all CS models corresponded to a neuron with a soma surface area of 1 cm$^2$. In practice, the size of the neuron would have a significant impact on the magnitudes of the input currents used. Larger cells will require larger values of input current to produce meaningful changes in the membrane voltage compared to smaller cells \cite{sterratt_principles_2011} and smaller cells would not survive larger injected currents. However, in a real patch clamp scenario the researcher would know a reasonable range of voltages that would produce neural firing. Given enough data, the DDF model would be able to learn the relationship between the magnitude of input needed to drive the neuron without needing an estimate of the soma size. 

\newpage
\section{Results}
\subsection{Homogeneous System Control}
Our first test of MPC for neural control was to force a CS neuron to reproduce a previously recorded voltage trajectory (the reference trajectory $V^{ref}$). We refer to this as homogeneous system control because each CS model was forced to track a reference trajectory it previously produced. For each CS model, we simulated 50 trials of 1-second responses to a chaotic current similar to the one used in the training data. We then used MPC with the corresponding DDF model to find $I_{inj}$ such that the errors in state tracking were minimized, thereby forcing the CS model to reproduce each of these 50 trials. 

We repeated each trial in an open-loop condition; that is, by injecting the same input current that produced the reference trajectory. If the neuron were deterministic, then the voltage trace would perfectly match the reference. However, because $I_{noise}$ varies in each trial, the same injected current will not produce the same voltage trace or pattern of spiking. Thus, the open-loop condition gives us a reference for the amount of variability we would expect to see without feedback control.

In order to quantify how well the MPC and open-loop control performed, we used three measures of fit: MSE, ISI-distance, and spike-distance. The MSE was calculated by comparing the reference trajectory $V^{ref}$ with the voltage trajectory $V$ the controlled CS model produced. The ISI- and spike-distances are measures of spike train similarity \cite{mulansky_pyspikepython_2016}. Rather than directly comparing the trajectories, these measures use the times that the CS model spiked and compare them to corresponding spikes in $V^{ref}$. Spike times were obtained by recording when the voltage exceeded a threshold (30 mV). The ISI-distance measures the similarity between the inter-spike intervals (ISI) of two spike trains, and the spike-distance measures the similarity of the spike timing between the two spike trains. 

\begin{figure}
    \includegraphics[width=1\textwidth]{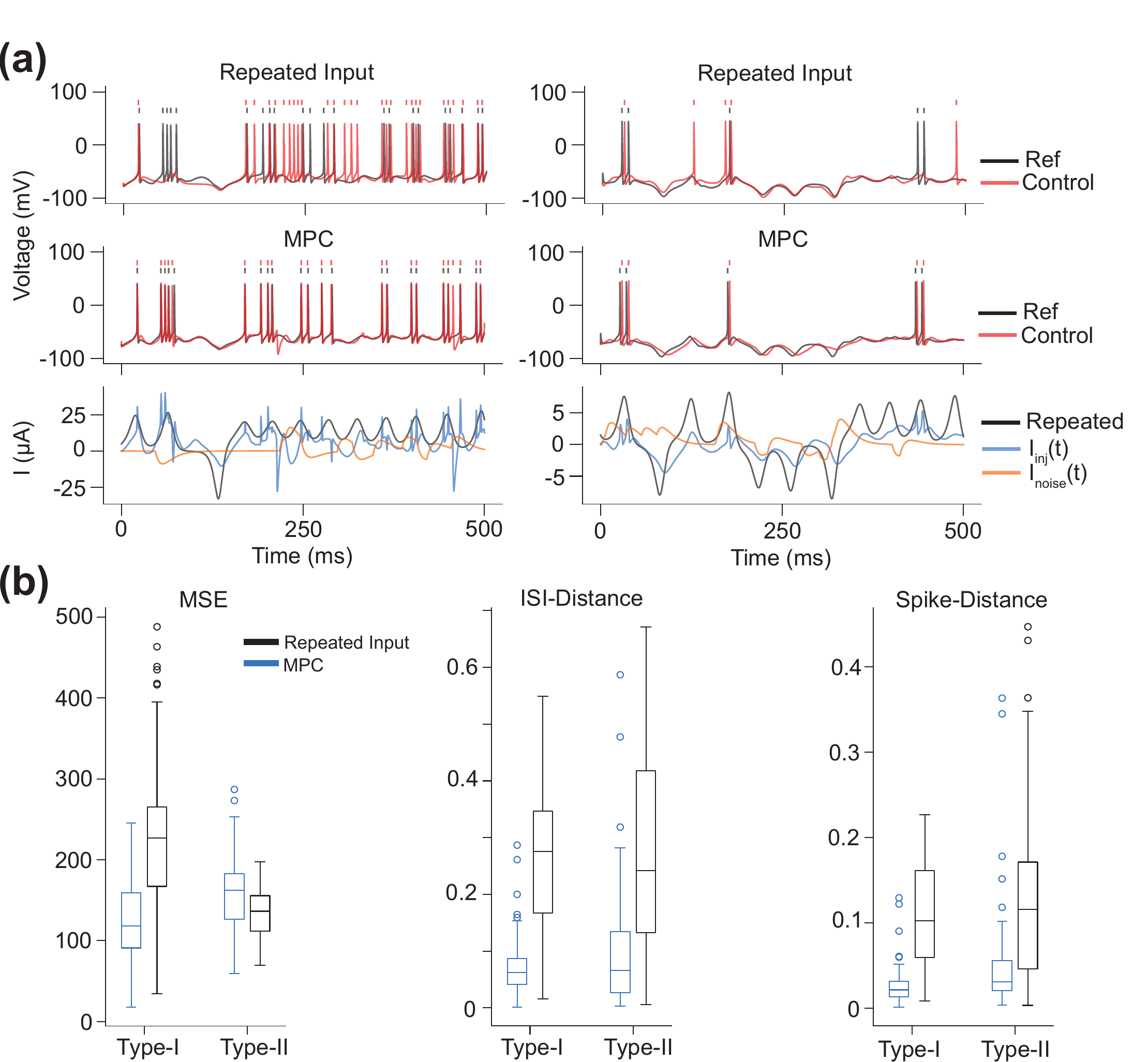}
    \caption{\textbf{Results of Homogeneous System Control.} (a) (Left) Example of a membrane voltage trajectory for a Type-I CS model with open-loop control. In black is the reference trajectory and in red is the controlled trajectory. Rasters above indicate spike times. The same $I_{inj}(t)$ used to generate the reference trajectory was used as the input for the open-loop control. However, the unknown noise current $I_{noise}(t)$ into the CS model resulted in the controlled membrane voltage deviating from the reference trajectory. (Middle) Example of a membrane voltage trajectory for a Type-I CS model controlled via MPC. The controlled membrane voltage tracks the reference trajectory extremely well compared to the open-loop controller. (Bottom) The unknown noise current (orange) into the CS model, the $I_{inj}(t)$ used to generate the reference trajectory and as the open-loop input (black), and MPC optimized input used to control the CS model (blue). (Right) The same diagrams as on the left, but for a Type-II CS model. (b) Performance metrics comparing the open-loop and MPC methods of control. The MSE was calculated for each reference/control trajectory pair. ISI-distance and spike-distance are both on the interval [0,1] where 0 indicates identical spike trains. Although the MSE for the Type-II CS model is slightly lower compared to the open-loop method, the spike train similarity metrics are much better for MPC.}
    \label{fig:Figure 6}
\end{figure}

As seen in Figure~\ref{fig:Figure 6}, MPC performed much better than open-loop control for both CS model types. This is clear from visual inspection of the membrane voltages and from the quantitative measures of performance. This result is remarkable because the underlying DDF model in the controller does not have any knowledge of the biophysical details of the system it is controlling.

\newpage

\subsection{Heterogeneous System Control}
As a more difficult test of the MPC controller, we investigated whether it could force a CS neuron of one type to follow a reference trajectory generated by the other type. In other words, could a Type-I neuron be made to spike like a Type-II neuron? We refer to this as heterogeneous system control because one dynamical system is being forced to behave as a different dynamical system. We used the same MSE and spike train similarity metrics to compare MPC performance with open-loop control. In this case, open-loop control was performed by taking the $I_{inj}$ that produced the $V^{ref}$ in a particular CS model type and using that as the command signal into the other CS model. 

\begin{figure}
    \centering
    \includegraphics[width=1\textwidth]{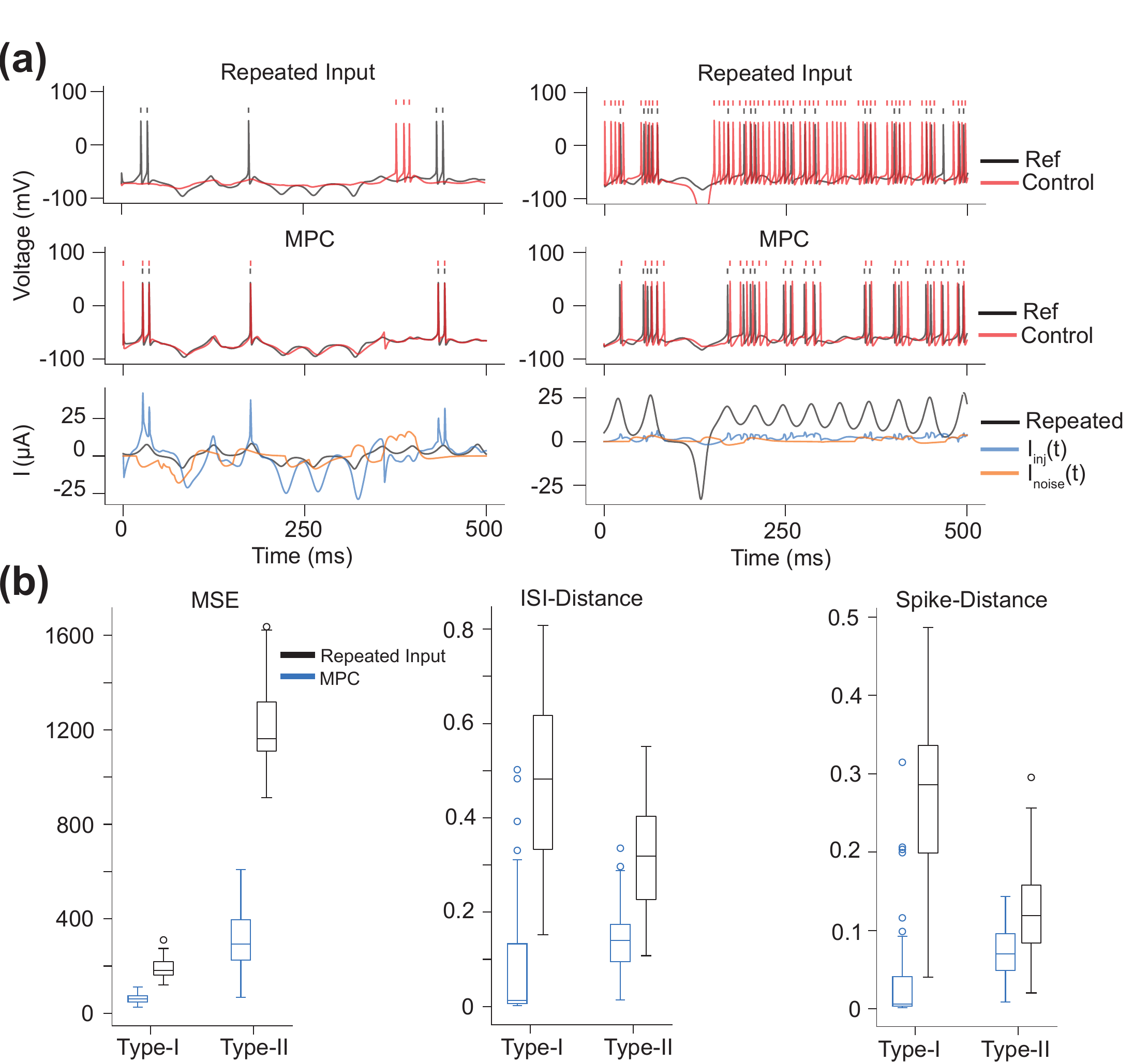}
    \caption{\textbf{Results of Heterogeneous System Control.} (a) (Left) Example of a membrane voltage trajectory for a Type-I CS model with open-loop control. The reference trajectory (black) was obtained from a Type-II CS model. In this example, the controlled trajectory (red) exhibited very little spiking and did not closely follow the reference trajectory. This was expected since the control input for the open-loop controller was the $I_{inj}(t)$ used to generate the reference trajectory. The Type-II model fires at a lower amplitude input than the Type-I model, and therefore the open-loop input would be weakly stimulating to the Type-I model. (Middle) Example of a membrane voltage trajectory for a Type-I CS model controlled via MPC. Notice that MPC is able to force the Type-I model to follow a Type-II model reference trajectory. (Bottom) The unknown noise current (orange) into the CS model, the $I_{inj}(t)$ used to generate the reference trajectory and as the open-loop input (black), and MPC optimized input used to control the CS model (blue). The MPC input drastically deviates from the open-loop control input in order to make the Type-I model follow the Type-II reference trajectory. (Right) The same diagrams as on the left, but now a Type-II CS model is controlled to follow a reference trajectory taken from a Type-I model. In open-loop control (Top), the Type-II model fires noticeably more than the reference trajectory since the control input is the $I_{inj}(t)$ used to generate the Type-I reference trajectory (which fires at higher input amplitudes compared to the Type-II model). However, MPC is able to more accurately control the Type-II neuron into following the Type-I reference trajectory. (b) Performance metrics comparing the open-loop and MPC methods of control. In all cases, MPC achieved much better control than the open-loop controller.}
    \label{fig:Figure 7}
\end{figure}

Unsurprisingly, open-loop control performed poorly for this task, because each CS model was a distinct dynamical system and thus responded differently to the same command signal. Type-I neurons often failed to fire at all when driven by injected currents used to stimulate Type-II neurons, presumably because the A-type current in the Type-I model counteracted the depolarizing injected current. Conversely, the Type-II neurons had a tendency to produce too many spikes when injected with currents used to stimulate Type-I neurons. Despite the dissimilar intrinsic currents and dynamical topologies of the two CS models, MPC was able to force each CS model to follow the trajectory of the other model as seen in Figure~\ref{fig:Figure 7}.

\newpage

\subsection{Spike Train Control and Comparison to Other Control Methods}
In many neuroscience studies, the experimenter wants to make a neuron spike at specific times without caring too much about the subthreshold activity. Spike trains are a point process, whereas the DDF models forecast a continuous variable. To overcome this mismatch in data structure, we extracted the mean spike waveform from the training data of each CS model and embedded it into a time series of constant value chosen to be below the threshold. The peaks of these embedded waveforms matched the spike times of the reference spike train. By doing so, we were able to convert any spike train into a reference trajectory with units of mV. 

As shown in Figure~\ref{fig:Figure 8}(a), MPC achieved good control of spike timing for both CS models. However, as previously stated, there are decades of work showing that controlling the firing of individual neurons is readily achievable with simple methods. With the increased use of data-driven methods for complex control schemes, it is important to consider when methods like MPC are more beneficial compared to traditional methods of control. To illustrate this, we also controlled the CS models with a proportional controller and open-loop pulse control. The proportional controller scaled the state error by a gain parameter of $K_p = -2$ for both CS models. While this controller does not explicitly use constraints to find the control signal, the maximum and minimum values were clamped to the same values as the MPC constraints. This simple closed-loop controller demonstrated exceptional performance (Figure~\ref{fig:Figure 8}(b)) which is unsurprising given that this is essentially the methodology used in voltage-clamp \cite{sherman2012axon}. Similarly, an entirely open-loop 2ms pulse is able to produce consistent firing to the desired spike train of both CS models (Figure~\ref{fig:Figure 8}(c)). While this would not prevent spikes produced by the unknown noise current, it is completely adequate for low noise systems. 

Although models of individual neurons are a good test bed for building data-driven models for use in MPC, we again stress that this methodology is not universally superior to simpler alternative methods. In practice, MPC is most useful in high dimensional systems and when there are many constraints \cite{rakovic_handbook_2019}. Given the ever increasing amount of data that can be recorded from neural activity (e.g. \cite{steinmetz2021neuropixels}), we have no doubt that data-driven MPC will become an important experimental tool in neuroscience.
\afterpage{%
\begin{figure}
    \centering
    \includegraphics[width=1\textwidth]{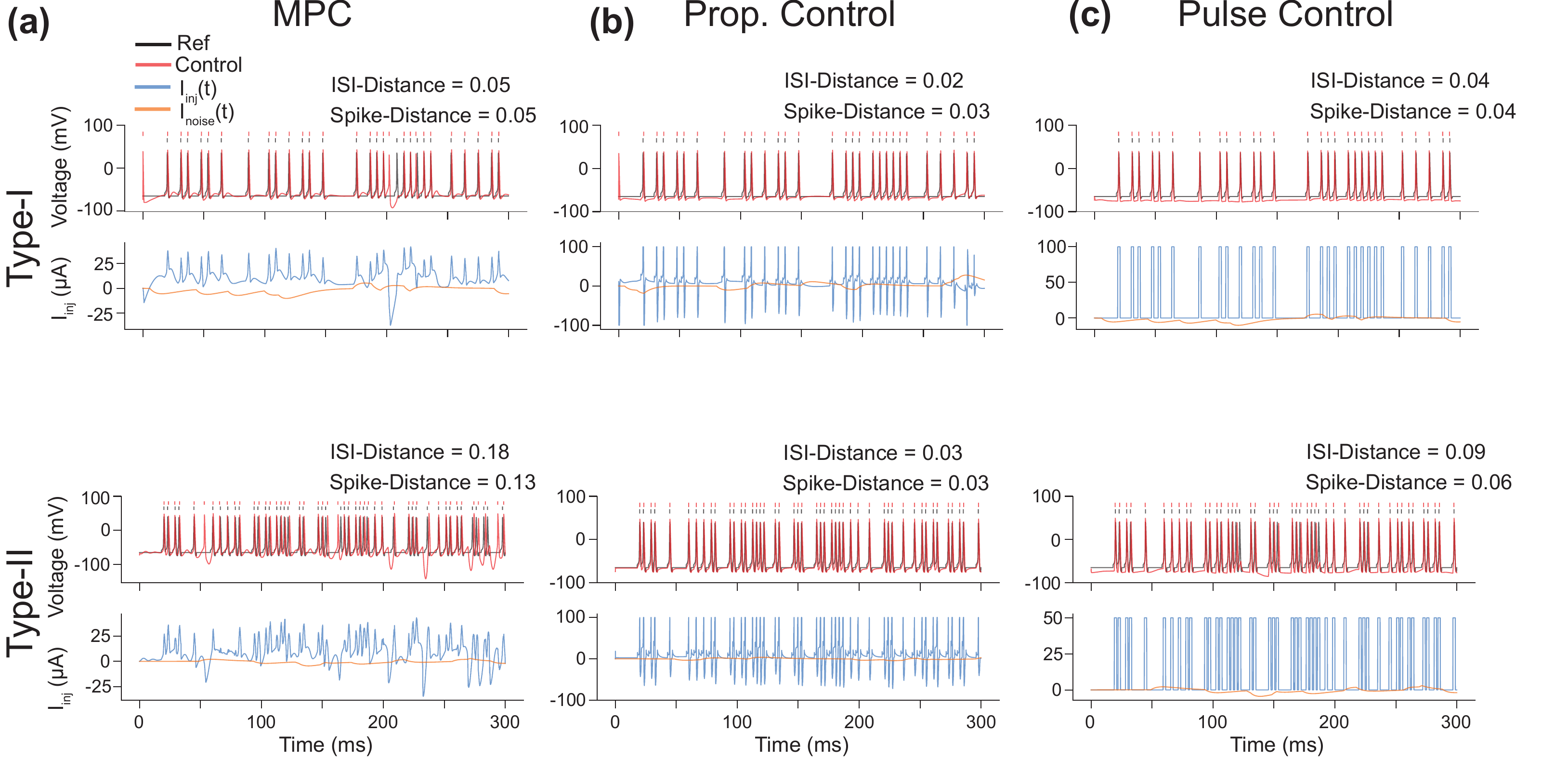}
    \caption{\textbf{Results of Spike Train Control.} (a) Membrane voltage trajectory controlled via MPC for a Type-I (top) and Type-II model (bottom) to follow a specific spike train. The reference trajectory (black) was obtained by taking the average spiking waveform for the corresponding CS model type and embedding it into a constant valued time series (-65 mV) at time points where a spike is desired. MPC was able to control CS models with their resulting membrane voltages (red) following the reference trajectories. (b) A proportional feedback controller is able to achieve much better performance compared to MPC in this example. The values of the control signal were clamped between $-100$ and $100$ $\mu A$ in accordance with the constraints but on the MPC controller. For both the Type-I and Type-II CS models, the proportional controller produces a sharp excitatory input at the desired spike time follwed by an strong inhibitory input. This inhibitory input is largely unneeed, but since the controller has no knowledge of the system dynamics (unlike MPC) it attempts to minimize the state error when the controlled spikes occur at a slightly delayed time compared to the reference spikes. (b) Open-loop control with $2$ ms wide pulses of current also produce spike trains exteremly similar to the desired set. The amplitudes of the two pulses required empirical scaling to ensure that the neurons would both fire when desired and not fire more than once to a single pulse.}
    \label{fig:Figure 8}
\end{figure}
\clearpage}
\newpage
\section{Discussion}
Neural systems can be difficult to control because of their nonlinear dynamics and many hidden states \cite{rabinovich_dynamical_2006}. Here we demonstrated that nonlinear MPC can control the well-characterized Connor-Stevens neuron model via an injected current using only measurements of the membrane voltage. The controller was able to force the model to reproduce a previously observed voltage trajectory in the presence of unknown intrinsic noise, follow a reference trajectory produced by a model with a different dynamical topology, and produce an arbitrary spike train with high temporal precision. Importantly, we are able to do this without any knowledge of the biophysical details of the Connor-Stevens model by using a data-driven forecasting model fit to a few seconds of noisy current-clamp data. 

This study represents one step toward the ultimate goal of controlling biological networks of neurons \emph{in vivo} to experimentally probe the mechanisms of neural computations and ameliorate pathological circuit states. The system tested here involves only a single neuron in the equivalent of a whole-cell patch recording, which enables an experimenter to make low-noise, high-bandwidth measurements of membrane potential while injecting current through an access resistance much smaller than the resting (input) resistance of the cell membrane. This preparation is easily controllable in practice however, and modern intracellular amplifiers are able to clamp cell voltage using relatively simple proportional feedback controllers implemented in analog circuitry \cite{sherman2012axon}. The purpose of this study was not to improve on amplifier design but rather as a proof of principle for how data-driven nonlinear MPC can achieve control of a neuron's membrane voltage without any prior knowledge of the intrinsic ionic currents expressed by a specific neuron. To our knowledge, all of the prior studies applying MPC to conductance-based neuron models have assumed knowledge about the neuron such as the states of current gating variables or the parameters and functional forms of their kinetics \cite{frohlich_feedback_2005,yue_non-linear_2022,ullah_tracking_2009,senthilvelmurugan_active_2023}, which would not be known in a real experiment. The results in this study show that this information is not necessary, bolstering confidence that data-driven MPC can be extended to neural networks in which there is likely to be even less knowledge about the full state and parameters of the system.

To illustrate some of the ways in which the principles in this study could be extended to networks, consider as an example the zebra finch's HVC, a bilateral premotor nucleus which contains around 36,000 highly interconnected neurons in each hemisphere \cite{bottjer_zebra_finch_hvc_1986}. Precise patterns of neural activity in HVC collectively result in the bird singing, and the ability to experimentally control HVC to produce arbitrary trajectories in its state space would produce major insights into how this system orchestrates vocal communication. Present technology allows simultaneous measurement of activity in a few hundred of these cells using calcium imaging or high-density extracellular electrophysiology, and activity could be manipulated in a (potentially different) subset of neurons using optogenetic stimulation. The state of the system would now be a vector, naively with one component for each of the neurons the experimenter was monitoring. The input would also be a vector corresponding to the neurons the experimenter was manipulating, and the loss function would generalize to a form along the lines of 
\begin{equation}
    J(x_0) = e_T^\intercal \mathbf{S} e_T + \sum_{n=0}^{T-1} e_{n+1}^\intercal \mathbf{Q} e_{n+1} + \Delta I_{n+1}^\intercal \mathbf{R} \Delta I_{n+1},
\end{equation}

where $e_i$ and $\Delta I_i$ have the same meaning as in equation \ref{eq:mpc_loss} but are now vector-valued with each element corresponding to an individual state and external input source respectively. The matrices $\mathbf{S}$, $\mathbf{Q}$, and $\mathbf{R}$ function largely the same as the scaling factors $\mathbf{s}$,$\mathbf{q}$, and $\mathbf{r}$ in equation \ref{eq:mpc_loss}, but now allow one to differentially weight the cost of each of the elements of vectors $e$ and $I$. For example, there may be a subset of neurons in a network that have an outsized impact on the population activity as a whole. By using larger values in the $\mathbf{Q}$ and $\mathbf{S}$ matrices that correspond to this subset of neurons in error vector $e$, the controller will view state errors in these neurons as more costly than the other units in the network. It should be noted that this kind of loss function could be applied across many different modalities of neural activity. Similar loss functions for linear MPC have been applied to optogenetics both \textit{in vivo} \cite{bolus_state-space_2021} and simulation models \cite{milias-argeitis_adaptive_2015,fox_bayesian_2023}. It would be straightforward (at least mathematically) to use behaviorally derived states in vector $e$ while maintaining cellular inputs in vector $I$. This would allow researchers to explore how specific patterns of neural activity control organism behavior.

Extending control to neural circuits may necessitate the use of more complicated function approximators to obtain a good forecasting model. However, there are several advantages of simpler models like the RBFN compared to more complex models. Time is often a constraint in neuroscience experiments and the ability to estimate and use a model in a short time frame is a necessity. When controlling a neuron (or neural circuit) with MPC, the forecasting model would need to be estimated quickly and in a data-efficient manner. Because neurons exhibit a large diversity of dynamics, a forecasting model trained on one neuron is unlikely to generalize to a different neuron. The more complex and time-consuming the forecasting model is to train, the less time there would be to control the neuron. We were able to construct DDF models using only 5 seconds of data and the training time was negligible compared to the amount of time for a typical patch-clamp experiment. Modern data-driven models often have many more parameters and more computationally expensive training algorithms which limit their ability to be practical in an experimental setting. Additionally, RBFNs can be estimated in an online setting (e.g., recursive least squares) which allow the DDF models to adapt to changes in the neural dynamics which can come from many sources such as electrode drift, tissue damage, and intrinsic plasticity. However, architectures such as RNNs and Transformers routinely achieve state of the art performance on time-series forecasting benchmarks and may be necessary when using MPC on large coupled networks.

Controlling networks may also require the inclusion of a state estimator in the control loop. In a whole-cell preparation, measurement noise is very low \cite{sherman2012axon} and one can assume that the measured values of the membrane voltage and injected current are the true values. In a neural circuit, it is not possible to achieve whole-cell access to more than a couple neurons at best, so it is necessary to use extracellular electrophysiology, which only reveals the timing of action potentials, or optical signals of calcium concentration, which tend to be slow and much noisier. Similarly, optogenetic stimulation of neural activity is much less precise than direct current injection. These sources of variability can corrupt the measurements of the system's state, leading to incorrect calculations of the optimal control inputs. If the structure of the noise can be assumed, techniques from robust MPC may lead to improved performance \cite{bemporad2007robust}. Robust MPC can provide a safer control scheme since the effect of disturbances is explicitly modeled and the controller tries to ensure the system does not enter a region of state space that is infeasible or potentially dangerous \cite{hewing_learning-based_2020}. State estimators can also transform the spike times arising from extracellular recording into estimates of a continuous latent state \cite{smith2010state}. Latent factor models could also be used as a latent state estimator which would reduce the dimensionality of the cost function thereby reducing the computational complexity of the optimization. Instead of using MPC to control the activity of every unit in the network, the factor model could express the coordinates of the network state and reference trajectory in the lower dimensional space \cite{fehrman2024model} (often referred to as the neural manifold \cite{langdon2023unifying}).

As our ability to collect vast quantities of neural data has surpassed our theoretical knowledge of the system dynamics, data-driven methods will be essential in increasing our ability to control the activity of the nervous system. By using nonlinear DDF models to approximate system dynamics, MPC has the potential to advance the field of neural control without needing a deep \textit{a priori} knowledge of the biophysics. This would allow for new experimental designs and reduce the need to have hand-engineered patterns of neural stimulation. Instead of the usual methods where an input is given to the system and the resulting behavior recorded, a complex and precise behavior could be defined in advance with the necessary stimulation found via MPC. There would be numerous therapeutic uses as well, with applications of MPC in neuroprosthetics being an especially promising area of research \cite{lambeth_robust_2023,wolf_trajectory_2022,singh_data-driven_2023,bao_model_2019}. Other therapeutic applications that would require network-level control include driving neural activity away from potentially pathological areas in state space (e.g., epilepsy \cite{chatterjee2020fractional,brar2018seizure}). Because MPC is an anticipatory controller, the dynamics model could forecast this activity before it happens and preemptively send a control signal to prevent the pathological state from occurring. The ability to have this level of control over therapeutic and experimental interventions will allow researchers to explore and validate new theories of neural dynamics as well as the relationship between extrinsic and intrinsic modulation of network activity.

\section{Appendix}
All source code used for this study can be found at \url{https://github.com/melizalab/mpc-hh}.
\subsection{DDF Model Derivation}
We assume that the continuous time dynamics of the membrane voltage are given by
\begin{equation}
    C\frac{dV}{dt} = F(V,X,\Theta,t)+I_{inj}.
\end{equation}
Let $t_n$ denote the $n$th time sample of measurement. Following \cite{clark_reduced-dimension_2022}, we can construct a DDF model by integrating over some time interval $\Delta t = t_{n+1} - t_n$,
\begin{equation}
    V_{n+1}-V_n = \frac{1}{C} \int_{t_n}^{t_{n+1}} F(V,X,\Theta,t')dt'+
    \frac{1}{C}\int_{t_n}^{t_{n+1}}I_{inj}(t')dt'
\end{equation}
where we denote $V(t_n)$ as $V_n$ for notational convenience. 

Since the injected current $I_{inj}$ is set by the researcher, it can be approximated by the trapezoidal rule,
\begin{equation}
    V_{n+1}-V_n = \frac{1}{C} \int_{t_n}^{t_{n+1}} F(V,X,\Theta,t')dt'+
    \alpha(I_{n+1}+I_n)
\end{equation}
where $\alpha = \frac{\Delta t}{2 C}$. We can now approximate the unknown integral with radial basis function network (RBFN) denoted by $F_{RBF}$,
\begin{equation}
    V_{n+1}= V_n+ F_{RBF}(S_n)+\alpha(I_{n+1}+I_n)
\end{equation}
where $S_n$ is a time-delay embedding of the membrane voltage. The time delay of each embedding $\tau$ and the number of delays $D_e$ are hyperparameters that must be tuned. For all simulations, we set $\tau^* = 1$ and $D_e = 2$, i.e. $S_n = (V_n,V_{n-1})$. The weights of the RBFN can be obtained by minimizing the cost function
\begin{equation}
    \sum_{i=0}^{N} \Bigl[ V_{i+1} - V_i - F_{RBF}(S_i)-\alpha I_{i+1}-\alpha I_i \Bigr]^2
\end{equation}
which can be obtained using standard least squares methods.
\subsection{Connor-Stevens Neuron}
\textbf{CS Model Equations}
\begin{align*}
C\frac{dV}{dt} & = I_{Na}+I_{K}+I_{A}+I_{l}+I_{noise}+I_{inj} \\
I_{Na} & = g_{Na} m^3 h (E_{Na}-V) \\ 
I_K & = g_K n^4 (E_K-V) \\
I_A & = g_A a^3 b (E_A-V) \\
I_l & = g_l(E_l-V)
\end{align*}

\textbf{CS Model Parameters}\\
\begin{center}
\begin{tabular}{ |c|c|c|c| } 
\hline
Parameter & Type-I (Type-II)\\
\hline
$C$ & 1 $\mu$Fcm\raise0.5ex\hbox{-2}\\
$E_{Na}$ & 50 mV \\ 
$E_K$ & -77 mV\\ 
$E_A$ & 80 mV  \\ 
$E_l$ & -22 (-72.8) mV \\ 
$g_{Na}$ & 120 mScm\raise0.5ex\hbox{-2} \\ 
$g_K$ & 20 mScm\raise0.5ex\hbox{-2}  \\ 
$g_A$ & 47.7 (0) mScm\raise0.5ex\hbox{-2}  \\ 
$g_l$ & 0.3 mScm\raise0.5ex\hbox{-2}  \\
\hline
\end{tabular}
\end{center}
The full list of parameter values and first-order kinetic equations can be found in \cite{sterratt_principles_2011}.
\subsection{MPC Hyperparameters}
\begin{table}[h!]
    \centering
    \setlength{\tabcolsep}{1.5em}
    \begin{tabular}{lcccc}
        \toprule
         & $\mathbf{q}$ & $\mathbf{s}$ & $\mathbf{r}$ & $T$\\
         \midrule
        Experiment I & & & & \\
        Type-I & 5 & 1 & 7 & 5\\
        Type-II & 1 & 0 & 100 & 5\\
        \bottomrule
        Experiment II & & & & \\
        Type-I & 5 & 1 & 7 & 5\\
        Type-II & 1 & 0 & 100 & 5\\
        \bottomrule
        Experiment III & & & & \\
        Type-I & 5 & 1 & 7 & 5\\
        Type-II & 5 & 1 & 50 & 5\\
        \bottomrule
    \end{tabular}
    \caption{MPC controller hyperparameters used for each experiment.}
    \label{tab:MPC_Hyperparameters}
\end{table}

\clearpage



\begin{thebibliography}{72}
\ifx \bisbn   \undefined \def \bisbn  #1{ISBN #1}\fi
\ifx \binits  \undefined \def \binits#1{#1}\fi
\ifx \bauthor  \undefined \def \bauthor#1{#1}\fi
\ifx \batitle  \undefined \def \batitle#1{#1}\fi
\ifx \bjtitle  \undefined \def \bjtitle#1{#1}\fi
\ifx \bvolume  \undefined \def \bvolume#1{\textbf{#1}}\fi
\ifx \byear  \undefined \def \byear#1{#1}\fi
\ifx \bissue  \undefined \def \bissue#1{#1}\fi
\ifx \bfpage  \undefined \def \bfpage#1{#1}\fi
\ifx \blpage  \undefined \def \blpage #1{#1}\fi
\ifx \burl  \undefined \def \burl#1{\textsf{#1}}\fi
\ifx \doiurl  \undefined \def \doiurl#1{\url{https://doi.org/#1}}\fi
\ifx \betal  \undefined \def \betal{\textit{et al.}}\fi
\ifx \binstitute  \undefined \def \binstitute#1{#1}\fi
\ifx \binstitutionaled  \undefined \def \binstitutionaled#1{#1}\fi
\ifx \bctitle  \undefined \def \bctitle#1{#1}\fi
\ifx \beditor  \undefined \def \beditor#1{#1}\fi
\ifx \bpublisher  \undefined \def \bpublisher#1{#1}\fi
\ifx \bbtitle  \undefined \def \bbtitle#1{#1}\fi
\ifx \bedition  \undefined \def \bedition#1{#1}\fi
\ifx \bseriesno  \undefined \def \bseriesno#1{#1}\fi
\ifx \blocation  \undefined \def \blocation#1{#1}\fi
\ifx \bsertitle  \undefined \def \bsertitle#1{#1}\fi
\ifx \bsnm \undefined \def \bsnm#1{#1}\fi
\ifx \bsuffix \undefined \def \bsuffix#1{#1}\fi
\ifx \bparticle \undefined \def \bparticle#1{#1}\fi
\ifx \barticle \undefined \def \barticle#1{#1}\fi
\bibcommenthead
\ifx \bconfdate \undefined \def \bconfdate #1{#1}\fi
\ifx \botherref \undefined \def \botherref #1{#1}\fi
\ifx \url \undefined \def \url#1{\textsf{#1}}\fi
\ifx \bchapter \undefined \def \bchapter#1{#1}\fi
\ifx \bbook \undefined \def \bbook#1{#1}\fi
\ifx \bcomment \undefined \def \bcomment#1{#1}\fi
\ifx \oauthor \undefined \def \oauthor#1{#1}\fi
\ifx \citeauthoryear \undefined \def \citeauthoryear#1{#1}\fi
\ifx \endbibitem  \undefined \def \endbibitem {}\fi
\ifx \bconflocation  \undefined \def \bconflocation#1{#1}\fi
\ifx \arxivurl  \undefined \def \arxivurl#1{\textsf{#1}}\fi
\csname PreBibitemsHook\endcsname

\bibitem[\protect\citeauthoryear{Schiff}{2011}]{schiff2011neural}
\begin{bbook}
\bauthor{\bsnm{Schiff}, \binits{S.J.}}:
\bbtitle{Neural Control Engineering: the Emerging Intersection Between Control
  Theory and Neuroscience}.
\bpublisher{MIT Press}, \blocation{Cambridge, MA}
(\byear{2011})
\end{bbook}
\endbibitem

\bibitem[\protect\citeauthoryear{Kao and Hennequin}{2019}]{kao2019neuroscience}
\begin{barticle}
\bauthor{\bsnm{Kao}, \binits{T.-C.}},
\bauthor{\bsnm{Hennequin}, \binits{G.}}:
\batitle{Neuroscience out of control: control-theoretic perspectives on neural
  circuit dynamics}.
\bjtitle{Current opinion in neurobiology}
\bvolume{58},
\bfpage{122}--\blpage{129}
(\byear{2019})
\end{barticle}
\endbibitem

\bibitem[\protect\citeauthoryear{Parkes et~al.}{2023}]{parkes2023using}
\begin{botherref}
\oauthor{\bsnm{Parkes}, \binits{L.}},
\oauthor{\bsnm{Kim}, \binits{J.Z.}},
\oauthor{\bsnm{Stiso}, \binits{J.}},
\oauthor{\bsnm{Brynildsen}, \binits{J.K.}},
\oauthor{\bsnm{Cieslak}, \binits{M.}},
\oauthor{\bsnm{Covitz}, \binits{S.}},
\oauthor{\bsnm{Gur}, \binits{R.E.}},
\oauthor{\bsnm{Gur}, \binits{R.C.}},
\oauthor{\bsnm{Pasqualetti}, \binits{F.}},
\oauthor{\bsnm{Shinohara}, \binits{R.T.}}, et al.:
Using network control theory to study the dynamics of the structural
  connectome.
bioRxiv
(2023)
\end{botherref}
\endbibitem

\bibitem[\protect\citeauthoryear{Emiliani
  et~al.}{2022}]{emiliani_optogenetics_2022}
\begin{barticle}
\bauthor{\bsnm{Emiliani}, \binits{V.}},
\bauthor{\bsnm{Entcheva}, \binits{E.}},
\bauthor{\bsnm{Hedrich}, \binits{R.}},
\bauthor{\bsnm{Hegemann}, \binits{P.}},
\bauthor{\bsnm{Konrad}, \binits{K.R.}},
\bauthor{\bsnm{Lüscher}, \binits{C.}},
\bauthor{\bsnm{Mahn}, \binits{M.}},
\bauthor{\bsnm{Pan}, \binits{Z.-H.}},
\bauthor{\bsnm{Sims}, \binits{R.R.}},
\bauthor{\bsnm{Vierock}, \binits{J.}},
\bauthor{\bsnm{Yizhar}, \binits{O.}}:
\batitle{Optogenetics for light control of biological systems}.
\bjtitle{Nature Reviews Methods Primers}
\bvolume{2}(\bissue{1}),
\bfpage{55}
(\byear{2022})
\doiurl{10.1038/s43586-022-00136-4}
\end{barticle}
\endbibitem

\bibitem[\protect\citeauthoryear{Grosenick
  et~al.}{2015}]{grosenick_closed-loop_2015}
\begin{barticle}
\bauthor{\bsnm{Grosenick}, \binits{L.}},
\bauthor{\bsnm{Marshel}, \binits{J.H.}},
\bauthor{\bsnm{Deisseroth}, \binits{K.}}:
\batitle{Closed-loop and activity-guided optogenetic control}.
\bjtitle{Neuron}
\bvolume{86}(\bissue{1}),
\bfpage{106}--\blpage{139}
(\byear{2015})
\doiurl{10.1016/j.neuron.2015.03.034}
\end{barticle}
\endbibitem

\bibitem[\protect\citeauthoryear{Zaaimi et~al.}{2022}]{zaaimi_closed-loop_2022}
\begin{barticle}
\bauthor{\bsnm{Zaaimi}, \binits{B.}},
\bauthor{\bsnm{Turnbull}, \binits{M.}},
\bauthor{\bsnm{Hazra}, \binits{A.}},
\bauthor{\bsnm{Wang}, \binits{Y.}},
\bauthor{\bsnm{Gandara}, \binits{C.}},
\bauthor{\bsnm{McLeod}, \binits{F.}},
\bauthor{\bsnm{McDermott}, \binits{E.E.}},
\bauthor{\bsnm{Escobedo-Cousin}, \binits{E.}},
\bauthor{\bsnm{Idil}, \binits{A.S.}},
\bauthor{\bsnm{Bailey}, \binits{R.G.}},
\bauthor{\bsnm{Tardio}, \binits{S.}},
\bauthor{\bsnm{Patel}, \binits{A.}},
\bauthor{\bsnm{Ponon}, \binits{N.}},
\bauthor{\bsnm{Gausden}, \binits{J.}},
\bauthor{\bsnm{Walsh}, \binits{D.}},
\bauthor{\bsnm{Hutchings}, \binits{F.}},
\bauthor{\bsnm{Kaiser}, \binits{M.}},
\bauthor{\bsnm{Cunningham}, \binits{M.O.}},
\bauthor{\bsnm{Clowry}, \binits{G.J.}},
\bauthor{\bsnm{LeBeau}, \binits{F.E.N.}},
\bauthor{\bsnm{Constandinou}, \binits{T.G.}},
\bauthor{\bsnm{Baker}, \binits{S.N.}},
\bauthor{\bsnm{Donaldson}, \binits{N.}},
\bauthor{\bsnm{Degenaar}, \binits{P.}},
\bauthor{\bsnm{O’Neill}, \binits{A.}},
\bauthor{\bsnm{Trevelyan}, \binits{A.J.}},
\bauthor{\bsnm{Jackson}, \binits{A.}}:
\batitle{Closed-loop optogenetic control of the dynamics of neural activity in
  non-human primates}.
\bjtitle{Nature Biomedical Engineering}
\bvolume{7}(\bissue{4}),
\bfpage{559}--\blpage{575}
(\byear{2022})
\doiurl{10.1038/s41551-022-00945-8}
\end{barticle}
\endbibitem

\bibitem[\protect\citeauthoryear{Nowotny and
  Levi}{2014}]{jaeger_voltage-clamp_2014}
\begin{bchapter}
\bauthor{\bsnm{Nowotny}, \binits{T.}},
\bauthor{\bsnm{Levi}, \binits{R.}}:
\bctitle{Voltage-{Clamp} {Technique}}.
In: \beditor{\bsnm{Jaeger}, \binits{D.}},
\beditor{\bsnm{Jung}, \binits{R.}} (eds.)
\bbtitle{Encyclopedia of {Computational} {Neuroscience}},
pp. \bfpage{1}--\blpage{5}.
\bpublisher{Springer},
\blocation{New York, NY}
(\byear{2014}).
\doiurl{10.1007/978-1-4614-7320-6_137-2}
\end{bchapter}
\endbibitem

\bibitem[\protect\citeauthoryear{Stefani}{2002}]{stefani_design_2002}
\begin{bbook}
\beditor{\bsnm{Stefani}, \binits{R.T.}} (ed.):
\bbtitle{Design of Feedback Control Systems},
\bedition{4th ed} edn.
\bsertitle{The {Oxford} series in electrical and computer engineering}.
\bpublisher{Oxford University Press},
\blocation{New York}
(\byear{2002})
\end{bbook}
\endbibitem

\bibitem[\protect\citeauthoryear{Shanechi et~al.}{2017}]{shanechi_rapid_2017}
\begin{barticle}
\bauthor{\bsnm{Shanechi}, \binits{M.M.}},
\bauthor{\bsnm{Orsborn}, \binits{A.L.}},
\bauthor{\bsnm{Moorman}, \binits{H.G.}},
\bauthor{\bsnm{Gowda}, \binits{S.}},
\bauthor{\bsnm{Dangi}, \binits{S.}},
\bauthor{\bsnm{Carmena}, \binits{J.M.}}:
\batitle{Rapid control and feedback rates enhance neuroprosthetic control}.
\bjtitle{Nature Communications}
\bvolume{8}(\bissue{1}),
\bfpage{13825}
(\byear{2017})
\doiurl{10.1038/ncomms13825}
\end{barticle}
\endbibitem

\bibitem[\protect\citeauthoryear{Gilja et~al.}{2012}]{gilja_brain_2012}
\begin{bchapter}
\bauthor{\bsnm{Gilja}, \binits{V.}},
\bauthor{\bsnm{Nuyujukian}, \binits{P.}},
\bauthor{\bsnm{Chestek}, \binits{C.A.}},
\bauthor{\bsnm{Cunningham}, \binits{J.P.}},
\bauthor{\bsnm{Yu}, \binits{B.M.}},
\bauthor{\bsnm{Fan}, \binits{J.M.}},
\bauthor{\bsnm{Ryu}, \binits{S.I.}},
\bauthor{\bsnm{Shenoy}, \binits{K.V.}}:
\bctitle{A brain machine interface control algorithm designed from a feedback
  control perspective}.
In: \bbtitle{2012 {Annual} {International} {Conference} of the {IEEE}
  {Engineering} in {Medicine} and {Biology} {Society}},
pp. \bfpage{1318}--\blpage{1322}.
\bpublisher{IEEE},
\blocation{San Diego, CA}
(\byear{2012}).
\doiurl{10.1109/EMBC.2012.6346180}
\end{bchapter}
\endbibitem

\bibitem[\protect\citeauthoryear{Willett et~al.}{2017}]{willett_feedback_2017}
\begin{barticle}
\bauthor{\bsnm{Willett}, \binits{F.R.}},
\bauthor{\bsnm{Pandarinath}, \binits{C.}},
\bauthor{\bsnm{Jarosiewicz}, \binits{B.}},
\bauthor{\bsnm{Murphy}, \binits{B.A.}},
\bauthor{\bsnm{Memberg}, \binits{W.D.}},
\bauthor{\bsnm{Blabe}, \binits{C.H.}},
\bauthor{\bsnm{Saab}, \binits{J.}},
\bauthor{\bsnm{Walter}, \binits{B.L.}},
\bauthor{\bsnm{Sweet}, \binits{J.A.}},
\bauthor{\bsnm{Miller}, \binits{J.P.}},
\bauthor{\bsnm{Henderson}, \binits{J.M.}},
\bauthor{\bsnm{Shenoy}, \binits{K.V.}},
\bauthor{\bsnm{Simeral}, \binits{J.D.}},
\bauthor{\bsnm{Hochberg}, \binits{L.R.}},
\bauthor{\bsnm{Kirsch}, \binits{R.F.}},
\bauthor{\bsnm{Ajiboye}, \binits{A.B.}}:
\batitle{Feedback control policies employed by people using intracortical
  brain–computer interfaces}.
\bjtitle{Journal of Neural Engineering}
\bvolume{14}(\bissue{1}),
\bfpage{016001}
(\byear{2017})
\doiurl{10.1088/1741-2560/14/1/016001}
\end{barticle}
\endbibitem

\bibitem[\protect\citeauthoryear{Zhang et~al.}{2021}]{zhang_prototype_2021}
\begin{barticle}
\bauthor{\bsnm{Zhang}, \binits{Q.}},
\bauthor{\bsnm{Hu}, \binits{S.}},
\bauthor{\bsnm{Talay}, \binits{R.}},
\bauthor{\bsnm{Xiao}, \binits{Z.}},
\bauthor{\bsnm{Rosenberg}, \binits{D.}},
\bauthor{\bsnm{Liu}, \binits{Y.}},
\bauthor{\bsnm{Sun}, \binits{G.}},
\bauthor{\bsnm{Li}, \binits{A.}},
\bauthor{\bsnm{Caravan}, \binits{B.}},
\bauthor{\bsnm{Singh}, \binits{A.}},
\bauthor{\bsnm{Gould}, \binits{J.D.}},
\bauthor{\bsnm{Chen}, \binits{Z.S.}},
\bauthor{\bsnm{Wang}, \binits{J.}}:
\batitle{A prototype closed-loop brain–machine interface for the study and
  treatment of pain}.
\bjtitle{Nature Biomedical Engineering}
\bvolume{7}(\bissue{4}),
\bfpage{533}--\blpage{545}
(\byear{2021})
\doiurl{10.1038/s41551-021-00736-7}
\end{barticle}
\endbibitem

\bibitem[\protect\citeauthoryear{Shanechi et~al.}{2016}]{shanechi_robust_2016}
\begin{barticle}
\bauthor{\bsnm{Shanechi}, \binits{M.M.}},
\bauthor{\bsnm{Orsborn}, \binits{A.L.}},
\bauthor{\bsnm{Carmena}, \binits{J.M.}}:
\batitle{Robust brain-machine interface design using optimal feedback control
  modeling and adaptive point process filtering}.
\bjtitle{PLOS Computational Biology}
\bvolume{12}(\bissue{4}),
\bfpage{1004730}
(\byear{2016})
\doiurl{10.1371/journal.pcbi.1004730}
\end{barticle}
\endbibitem

\bibitem[\protect\citeauthoryear{Cunningham
  et~al.}{2011}]{cunningham_closed-loop_2011}
\begin{barticle}
\bauthor{\bsnm{Cunningham}, \binits{J.P.}},
\bauthor{\bsnm{Nuyujukian}, \binits{P.}},
\bauthor{\bsnm{Gilja}, \binits{V.}},
\bauthor{\bsnm{Chestek}, \binits{C.A.}},
\bauthor{\bsnm{Ryu}, \binits{S.I.}},
\bauthor{\bsnm{Shenoy}, \binits{K.V.}}:
\batitle{A closed-loop human simulator for investigating the role of feedback
  control in brain-machine interfaces}.
\bjtitle{Journal of Neurophysiology}
\bvolume{105}(\bissue{4}),
\bfpage{1932}--\blpage{1949}
(\byear{2011})
\doiurl{10.1152/jn.00503.2010}
\end{barticle}
\endbibitem

\bibitem[\protect\citeauthoryear{Wright et~al.}{2016}]{wright_review_2016}
\begin{botherref}
\oauthor{\bsnm{Wright}, \binits{J.}},
\oauthor{\bsnm{Macefield}, \binits{V.G.}},
\oauthor{\bsnm{Van~Schaik}, \binits{A.}},
\oauthor{\bsnm{Tapson}, \binits{J.C.}}:
A review of control strategies in closed-loop neuroprosthetic systems.
Frontiers in Neuroscience
\textbf{10}
(2016)
\doiurl{10.3389/fnins.2016.00312}
\end{botherref}
\endbibitem

\bibitem[\protect\citeauthoryear{Pandarinath and
  Bensmaia}{2022}]{pandarinath_science_2022}
\begin{barticle}
\bauthor{\bsnm{Pandarinath}, \binits{C.}},
\bauthor{\bsnm{Bensmaia}, \binits{S.J.}}:
\batitle{The science and engineering behind sensitized brain-controlled bionic
  hands}.
\bjtitle{Physiological Reviews}
\bvolume{102}(\bissue{2}),
\bfpage{551}--\blpage{604}
(\byear{2022})
\doiurl{10.1152/physrev.00034.2020}
\end{barticle}
\endbibitem

\bibitem[\protect\citeauthoryear{Pedrocchi et~al.}{2006}]{pedrocchi_error_2006}
\begin{barticle}
\bauthor{\bsnm{Pedrocchi}, \binits{A.}},
\bauthor{\bsnm{Ferrante}, \binits{S.}},
\bauthor{\bsnm{De~Momi}, \binits{E.}},
\bauthor{\bsnm{Ferrigno}, \binits{G.}}:
\batitle{Error mapping controller: a closed loop neuroprosthesis controlled by
  artificial neural networks}.
\bjtitle{Journal of NeuroEngineering and Rehabilitation}
\bvolume{3}(\bissue{1}),
\bfpage{25}
(\byear{2006})
\doiurl{10.1186/1743-0003-3-25}
\end{barticle}
\endbibitem

\bibitem[\protect\citeauthoryear{Bolus et~al.}{2021}]{bolus_state-space_2021}
\begin{barticle}
\bauthor{\bsnm{Bolus}, \binits{M.F.}},
\bauthor{\bsnm{Willats}, \binits{A.A.}},
\bauthor{\bsnm{Rozell}, \binits{C.J.}},
\bauthor{\bsnm{Stanley}, \binits{G.B.}}:
\batitle{State-space optimal feedback control of optogenetically driven neural
  activity}.
\bjtitle{Journal of Neural Engineering}
\bvolume{18}(\bissue{3}),
\bfpage{036006}
(\byear{2021})
\doiurl{10.1088/1741-2552/abb89c}
\end{barticle}
\endbibitem

\bibitem[\protect\citeauthoryear{Bergs et~al.}{2023}]{bergs_all-optical_2023}
\begin{barticle}
\bauthor{\bsnm{Bergs}, \binits{A.C.F.}},
\bauthor{\bsnm{Liewald}, \binits{J.F.}},
\bauthor{\bsnm{Rodriguez-Rozada}, \binits{S.}},
\bauthor{\bsnm{Liu}, \binits{Q.}},
\bauthor{\bsnm{Wirt}, \binits{C.}},
\bauthor{\bsnm{Bessel}, \binits{A.}},
\bauthor{\bsnm{Zeitzschel}, \binits{N.}},
\bauthor{\bsnm{Durmaz}, \binits{H.}},
\bauthor{\bsnm{Nozownik}, \binits{A.}},
\bauthor{\bsnm{Dill}, \binits{H.}},
\bauthor{\bsnm{Jospin}, \binits{M.}},
\bauthor{\bsnm{Vierock}, \binits{J.}},
\bauthor{\bsnm{Bargmann}, \binits{C.I.}},
\bauthor{\bsnm{Hegemann}, \binits{P.}},
\bauthor{\bsnm{Wiegert}, \binits{J.S.}},
\bauthor{\bsnm{Gottschalk}, \binits{A.}}:
\batitle{All-optical closed-loop voltage clamp for precise control of muscles
  and neurons in live animals}.
\bjtitle{Nature Communications}
\bvolume{14}(\bissue{1}),
\bfpage{1939}
(\byear{2023})
\doiurl{10.1038/s41467-023-37622-6}
\end{barticle}
\endbibitem

\bibitem[\protect\citeauthoryear{Newman et~al.}{2015}]{newman_optogenetic_2015}
\begin{barticle}
\bauthor{\bsnm{Newman}, \binits{J.P.}},
\bauthor{\bsnm{Fong}, \binits{M.-f.}},
\bauthor{\bsnm{Millard}, \binits{D.C.}},
\bauthor{\bsnm{Whitmire}, \binits{C.J.}},
\bauthor{\bsnm{Stanley}, \binits{G.B.}},
\bauthor{\bsnm{Potter}, \binits{S.M.}}:
\batitle{Optogenetic feedback control of neural activity}.
\bjtitle{eLife}
\bvolume{4},
\bfpage{07192}
(\byear{2015})
\doiurl{10.7554/eLife.07192}
\end{barticle}
\endbibitem

\bibitem[\protect\citeauthoryear{Hewing
  et~al.}{2020}]{hewing_learning-based_2020}
\begin{barticle}
\bauthor{\bsnm{Hewing}, \binits{L.}},
\bauthor{\bsnm{Wabersich}, \binits{K.P.}},
\bauthor{\bsnm{Menner}, \binits{M.}},
\bauthor{\bsnm{Zeilinger}, \binits{M.N.}}:
\batitle{Learning-based model predictive control: Toward safe learning in
  control}.
\bjtitle{Annual Review of Control, Robotics, and Autonomous Systems}
\bvolume{3}(\bissue{1}),
\bfpage{269}--\blpage{296}
(\byear{2020})
\doiurl{10.1146/annurev-control-090419-075625}
\end{barticle}
\endbibitem

\bibitem[\protect\citeauthoryear{Holkar and
  Waghmare}{2010}]{holkar2010overview}
\begin{barticle}
\bauthor{\bsnm{Holkar}, \binits{K.}},
\bauthor{\bsnm{Waghmare}, \binits{L.M.}}:
\batitle{An overview of model predictive control}.
\bjtitle{International Journal of Control and Automation}
\bvolume{3}(\bissue{4}),
\bfpage{47}--\blpage{63}
(\byear{2010})
\end{barticle}
\endbibitem

\bibitem[\protect\citeauthoryear{Lin et~al.}{2023}]{lin_development_2023}
\begin{barticle}
\bauthor{\bsnm{Lin}, \binits{L.}},
\bauthor{\bsnm{Oncken}, \binits{J.}},
\bauthor{\bsnm{Agarwal}, \binits{V.}},
\bauthor{\bsnm{Permann}, \binits{C.}},
\bauthor{\bsnm{Gribok}, \binits{A.}},
\bauthor{\bsnm{McJunkin}, \binits{T.}},
\bauthor{\bsnm{Eggers}, \binits{S.}},
\bauthor{\bsnm{Boring}, \binits{R.}}:
\batitle{Development and assessment of a model predictive controller enabling
  anticipatory control strategies for a heat-pipe system}.
\bjtitle{Progress in Nuclear Energy}
\bvolume{156},
\bfpage{104527}
(\byear{2023})
\doiurl{10.1016/j.pnucene.2022.104527}
\end{barticle}
\endbibitem

\bibitem[\protect\citeauthoryear{Raković and
  Levine}{2019}]{rakovic_handbook_2019}
\begin{bbook}
\beditor{\bsnm{Raković}, \binits{S.V.}},
\beditor{\bsnm{Levine}, \binits{W.S.}} (eds.):
\bbtitle{Handbook of {Model} {Predictive} {Control}}.
\bsertitle{Control {Engineering}}.
\bpublisher{Springer},
\blocation{Cham}
(\byear{2019}).
\doiurl{10.1007/978-3-319-77489-3}
\end{bbook}
\endbibitem

\bibitem[\protect\citeauthoryear{Brunton and
  Kutz}{2019}]{brunton_data-driven_2019}
\begin{bbook}
\bauthor{\bsnm{Brunton}, \binits{S.L.}},
\bauthor{\bsnm{Kutz}, \binits{J.N.}}:
\bbtitle{Data-driven Science and Engineering: Machine Learning, Dynamical
  Systems, and Control}.
\bpublisher{Cambridge University Press},
\blocation{Cambridge}
(\byear{2019})
\end{bbook}
\endbibitem

\bibitem[\protect\citeauthoryear{Schwenzer
  et~al.}{2021}]{schwenzer_review_2021}
\begin{barticle}
\bauthor{\bsnm{Schwenzer}, \binits{M.}},
\bauthor{\bsnm{Ay}, \binits{M.}},
\bauthor{\bsnm{Bergs}, \binits{T.}},
\bauthor{\bsnm{Abel}, \binits{D.}}:
\batitle{Review on model predictive control: an engineering perspective}.
\bjtitle{The International Journal of Advanced Manufacturing Technology}
\bvolume{117}(\bissue{5-6}),
\bfpage{1327}--\blpage{1349}
(\byear{2021})
\doiurl{10.1007/s00170-021-07682-3}
\end{barticle}
\endbibitem

\bibitem[\protect\citeauthoryear{Rabinovich
  et~al.}{2006}]{rabinovich_dynamical_2006}
\begin{barticle}
\bauthor{\bsnm{Rabinovich}, \binits{M.I.}},
\bauthor{\bsnm{Varona}, \binits{P.}},
\bauthor{\bsnm{Selverston}, \binits{A.I.}},
\bauthor{\bsnm{Abarbanel}, \binits{H.D.I.}}:
\batitle{Dynamical principles in neuroscience}.
\bjtitle{Reviews of Modern Physics}
\bvolume{78}(\bissue{4}),
\bfpage{1213}--\blpage{1265}
(\byear{2006})
\doiurl{10.1103/RevModPhys.78.1213}
\end{barticle}
\endbibitem

\bibitem[\protect\citeauthoryear{Toth et~al.}{2011}]{toth_dynamical_2011}
\begin{barticle}
\bauthor{\bsnm{Toth}, \binits{B.A.}},
\bauthor{\bsnm{Kostuk}, \binits{M.}},
\bauthor{\bsnm{Meliza}, \binits{C.D.}},
\bauthor{\bsnm{Margoliash}, \binits{D.}},
\bauthor{\bsnm{Abarbanel}, \binits{H.D.I.}}:
\batitle{Dynamical estimation of neuron and network properties {I}: variational
  methods}.
\bjtitle{Biological Cybernetics}
\bvolume{105}(\bissue{3-4}),
\bfpage{217}--\blpage{237}
(\byear{2011})
\doiurl{10.1007/s00422-011-0459-1}
\end{barticle}
\endbibitem

\bibitem[\protect\citeauthoryear{Fröhlich and
  Jezernik}{2005}]{frohlich_feedback_2005}
\begin{barticle}
\bauthor{\bsnm{Fröhlich}, \binits{F.}},
\bauthor{\bsnm{Jezernik}, \binits{S.}}:
\batitle{Feedback control of {Hodgkin}–{Huxley} nerve cell dynamics}.
\bjtitle{Control Engineering Practice}
\bvolume{13}(\bissue{9}),
\bfpage{1195}--\blpage{1206}
(\byear{2005})
\doiurl{10.1016/j.conengprac.2004.10.008}
\end{barticle}
\endbibitem

\bibitem[\protect\citeauthoryear{Yue et~al.}{2022}]{yue_non-linear_2022}
\begin{botherref}
\oauthor{\bsnm{Yue}, \binits{R.}},
\oauthor{\bsnm{Tomastik}, \binits{R.}},
\oauthor{\bsnm{Dutta}, \binits{A.}}:
Non-linear model-based control of neural cell dynamics.
preprint,
In Review
(May 2022).
\doiurl{10.21203/rs.3.rs-580874/v2}
\end{botherref}
\endbibitem

\bibitem[\protect\citeauthoryear{Senthilvelmurugan and
  Subbian}{2023}]{senthilvelmurugan_active_2023}
\begin{barticle}
\bauthor{\bsnm{Senthilvelmurugan}, \binits{N.N.}},
\bauthor{\bsnm{Subbian}, \binits{S.}}:
\batitle{Active fault tolerant deep brain stimulator for epilepsy using deep
  neural network}.
\bjtitle{Biomedical Engineering / Biomedizinische Technik}
\bvolume{68}(\bissue{4}),
\bfpage{373}--\blpage{392}
(\byear{2023})
\doiurl{10.1515/bmt-2021-0302}
\end{barticle}
\endbibitem

\bibitem[\protect\citeauthoryear{Kostuk et~al.}{2012}]{kostuk_dynamical_2012}
\begin{barticle}
\bauthor{\bsnm{Kostuk}, \binits{M.}},
\bauthor{\bsnm{Toth}, \binits{B.A.}},
\bauthor{\bsnm{Meliza}, \binits{C.D.}},
\bauthor{\bsnm{Margoliash}, \binits{D.}},
\bauthor{\bsnm{Abarbanel}, \binits{H.D.I.}}:
\batitle{Dynamical estimation of neuron and network properties {II}: path
  integral {Monte} {Carlo} methods}.
\bjtitle{Biological Cybernetics}
\bvolume{106}(\bissue{3}),
\bfpage{155}--\blpage{167}
(\byear{2012})
\doiurl{10.1007/s00422-012-0487-5}
\end{barticle}
\endbibitem

\bibitem[\protect\citeauthoryear{Ullah and Schiff}{2009}]{ullah_tracking_2009}
\begin{barticle}
\bauthor{\bsnm{Ullah}, \binits{G.}},
\bauthor{\bsnm{Schiff}, \binits{S.J.}}:
\batitle{Tracking and control of neuronal {Hodgkin}-{Huxley} dynamics}.
\bjtitle{Physical Review E}
\bvolume{79}(\bissue{4}),
\bfpage{040901}
(\byear{2009})
\doiurl{10.1103/PhysRevE.79.040901}
\end{barticle}
\endbibitem

\bibitem[\protect\citeauthoryear{Meliza et~al.}{2014}]{Meliza:2014p15146}
\begin{barticle}
\bauthor{\bsnm{Meliza}, \binits{C.D.}},
\bauthor{\bsnm{Kostuk}, \binits{M.}},
\bauthor{\bsnm{Huang}, \binits{H.}},
\bauthor{\bsnm{Nogaret}, \binits{A.}},
\bauthor{\bsnm{Margoliash}, \binits{D.}},
\bauthor{\bsnm{Abarbanel}, \binits{H.D.I.}}:
\batitle{{Estimating parameters and predicting membrane voltages with
  conductance-based neuron models.}}
\bjtitle{Biol Cybern}
\bvolume{108}(\bissue{4}),
\bfpage{495}--\blpage{516}
(\byear{2014})
\doiurl{10.1007/s00422-014-0615-5}
\end{barticle}
\endbibitem

\bibitem[\protect\citeauthoryear{Bieker et~al.}{2019}]{bieker_deep_2019}
\begin{botherref}
\oauthor{\bsnm{Bieker}, \binits{K.}},
\oauthor{\bsnm{Peitz}, \binits{S.}},
\oauthor{\bsnm{Brunton}, \binits{S.L.}},
\oauthor{\bsnm{Kutz}, \binits{J.N.}},
\oauthor{\bsnm{Dellnitz}, \binits{M.}}:
Deep model predictive control with online learning for complex physical systems
(2019)
\doiurl{10.48550/ARXIV.1905.10094} .
Publisher: arXiv Version Number: 1
\end{botherref}
\endbibitem

\bibitem[\protect\citeauthoryear{Kaiser et~al.}{2018}]{kaiser_sparse_2018}
\begin{barticle}
\bauthor{\bsnm{Kaiser}, \binits{E.}},
\bauthor{\bsnm{Kutz}, \binits{J.N.}},
\bauthor{\bsnm{Brunton}, \binits{S.L.}}:
\batitle{Sparse identification of nonlinear dynamics for model predictive
  control in the low-data limit}.
\bjtitle{Proceedings of the Royal Society A: Mathematical, Physical and
  Engineering Sciences}
\bvolume{474}(\bissue{2219}),
\bfpage{20180335}
(\byear{2018})
\doiurl{10.1098/rspa.2018.0335}
\end{barticle}
\endbibitem

\bibitem[\protect\citeauthoryear{Salzmann
  et~al.}{2023}]{salzmann_real-time_2023}
\begin{barticle}
\bauthor{\bsnm{Salzmann}, \binits{T.}},
\bauthor{\bsnm{Kaufmann}, \binits{E.}},
\bauthor{\bsnm{Arrizabalaga}, \binits{J.}},
\bauthor{\bsnm{Pavone}, \binits{M.}},
\bauthor{\bsnm{Scaramuzza}, \binits{D.}},
\bauthor{\bsnm{Ryll}, \binits{M.}}:
\batitle{Real-time neural {MPC}: Deep learning model predictive control for
  quadrotors and agile robotic platforms}.
\bjtitle{IEEE Robotics and Automation Letters}
\bvolume{8}(\bissue{4}),
\bfpage{2397}--\blpage{2404}
(\byear{2023})
\doiurl{10.1109/LRA.2023.3246839}
\end{barticle}
\endbibitem

\bibitem[\protect\citeauthoryear{Zheng and
  Wu}{2023}]{zheng_physics-informed_2023}
\begin{barticle}
\bauthor{\bsnm{Zheng}, \binits{Y.}},
\bauthor{\bsnm{Wu}, \binits{Z.}}:
\batitle{Physics-informed online machine learning and predictive control of
  nonlinear processes with parameter uncertainty}.
\bjtitle{Industrial \& Engineering Chemistry Research}
\bvolume{62}(\bissue{6}),
\bfpage{2804}--\blpage{2818}
(\byear{2023})
\doiurl{10.1021/acs.iecr.2c03691}
\end{barticle}
\endbibitem

\bibitem[\protect\citeauthoryear{Plaster and
  Kumar}{2019}]{plaster_data-driven_2019}
\begin{barticle}
\bauthor{\bsnm{Plaster}, \binits{B.}},
\bauthor{\bsnm{Kumar}, \binits{G.}}:
\batitle{Data-driven predictive modeling of neuronal dynamics using long
  short-term memory}.
\bjtitle{Algorithms}
\bvolume{12}(\bissue{10}),
\bfpage{203}
(\byear{2019})
\doiurl{10.3390/a12100203}
\end{barticle}
\endbibitem

\bibitem[\protect\citeauthoryear{Sherman-Gold}{2012}]{sherman2012axon}
\begin{botherref}
\oauthor{\bsnm{Sherman-Gold}, \binits{R.}}:
The axon guide, a guide to electrophysiology and biophysics laboratory
  techniques.
San Jose: Molecular Devices, LLC
(2012)
\end{botherref}
\endbibitem

\bibitem[\protect\citeauthoryear{Sterratt}{2011}]{sterratt_principles_2011}
\begin{bbook}
\beditor{\bsnm{Sterratt}, \binits{D.}} (ed.):
\bbtitle{Principles of Computational Modelling in Neuroscience}.
\bpublisher{Cambridge University Press},
\blocation{Cambridge ; New York}
(\byear{2011}).
\bcomment{OCLC: ocn690090171}
\end{bbook}
\endbibitem

\bibitem[\protect\citeauthoryear{Knowlton
  et~al.}{2014}]{knowlton_dynamical_2014}
\begin{barticle}
\bauthor{\bsnm{Knowlton}, \binits{C.}},
\bauthor{\bsnm{Meliza}, \binits{C.D.}},
\bauthor{\bsnm{Margoliash}, \binits{D.}},
\bauthor{\bsnm{Abarbanel}, \binits{H.D.I.}}:
\batitle{Dynamical estimation of neuron and network properties {III}: network
  analysis using neuron spike times}.
\bjtitle{Biological Cybernetics}
\bvolume{108}(\bissue{3}),
\bfpage{261}--\blpage{273}
(\byear{2014})
\doiurl{10.1007/s00422-014-0601-y}
\end{barticle}
\endbibitem

\bibitem[\protect\citeauthoryear{Skinner}{2006}]{skinner2006conductance}
\begin{barticle}
\bauthor{\bsnm{Skinner}, \binits{F.K.}}:
\batitle{Conductance-based models}.
\bjtitle{Scholarpedia}
\bvolume{1}(\bissue{11}),
\bfpage{1408}
(\byear{2006})
\end{barticle}
\endbibitem

\bibitem[\protect\citeauthoryear{Bourdeau
  et~al.}{2019}]{bourdeau_modeling_2019}
\begin{barticle}
\bauthor{\bsnm{Bourdeau}, \binits{M.}},
\bauthor{\bsnm{Zhai}, \binits{X.Q.}},
\bauthor{\bsnm{Nefzaoui}, \binits{E.}},
\bauthor{\bsnm{Guo}, \binits{X.}},
\bauthor{\bsnm{Chatellier}, \binits{P.}}:
\batitle{Modeling and forecasting building energy consumption: {A} review of
  data-driven techniques}.
\bjtitle{Sustainable Cities and Society}
\bvolume{48},
\bfpage{101533}
(\byear{2019})
\doiurl{10.1016/j.scs.2019.101533}
\end{barticle}
\endbibitem

\bibitem[\protect\citeauthoryear{Clark
  et~al.}{2022}]{clark_reduced-dimension_2022}
\begin{barticle}
\bauthor{\bsnm{Clark}, \binits{R.}},
\bauthor{\bsnm{Fuller}, \binits{L.}},
\bauthor{\bsnm{Platt}, \binits{J.A.}},
\bauthor{\bsnm{Abarbanel}, \binits{H.D.I.}}:
\batitle{Reduced-dimension, biophysical neuron models constructed from observed
  data}.
\bjtitle{Neural Computation}
\bvolume{34}(\bissue{7}),
\bfpage{1545}--\blpage{1587}
(\byear{2022})
\doiurl{10.1162/neco_a_01515}
\end{barticle}
\endbibitem

\bibitem[\protect\citeauthoryear{Lowe and
  Broomhead}{1988}]{lowe1988multivariable}
\begin{barticle}
\bauthor{\bsnm{Lowe}, \binits{D.}},
\bauthor{\bsnm{Broomhead}, \binits{D.}}:
\batitle{Multivariable functional interpolation and adaptive networks}.
\bjtitle{Complex systems}
\bvolume{2}(\bissue{3}),
\bfpage{321}--\blpage{355}
(\byear{1988})
\end{barticle}
\endbibitem

\bibitem[\protect\citeauthoryear{Hochreiter and
  Schmidhuber}{1997}]{hochreiter1997long}
\begin{barticle}
\bauthor{\bsnm{Hochreiter}, \binits{S.}},
\bauthor{\bsnm{Schmidhuber}, \binits{J.}}:
\batitle{Long short-term memory}.
\bjtitle{Neural computation}
\bvolume{9}(\bissue{8}),
\bfpage{1735}--\blpage{1780}
(\byear{1997})
\end{barticle}
\endbibitem

\bibitem[\protect\citeauthoryear{Vaswani et~al.}{2017}]{vaswani2017attention}
\begin{botherref}
\oauthor{\bsnm{Vaswani}, \binits{A.}},
\oauthor{\bsnm{Shazeer}, \binits{N.}},
\oauthor{\bsnm{Parmar}, \binits{N.}},
\oauthor{\bsnm{Uszkoreit}, \binits{J.}},
\oauthor{\bsnm{Jones}, \binits{L.}},
\oauthor{\bsnm{Gomez}, \binits{A.N.}},
\oauthor{\bsnm{Kaiser}, \binits{{\L}.}},
\oauthor{\bsnm{Polosukhin}, \binits{I.}}:
Attention is all you need.
Advances in neural information processing systems
\textbf{30}
(2017)
\end{botherref}
\endbibitem

\bibitem[\protect\citeauthoryear{Park and Sandberg}{1991}]{park_universal_1991}
\begin{barticle}
\bauthor{\bsnm{Park}, \binits{J.}},
\bauthor{\bsnm{Sandberg}, \binits{I.W.}}:
\batitle{Universal approximation using radial-basis-function networks}.
\bjtitle{Neural Computation}
\bvolume{3}(\bissue{2}),
\bfpage{246}--\blpage{257}
(\byear{1991})
\doiurl{10.1162/neco.1991.3.2.246}
\end{barticle}
\endbibitem

\bibitem[\protect\citeauthoryear{Clark et~al.}{2022}]{clark_data_2022}
\begin{botherref}
\oauthor{\bsnm{Clark}, \binits{R.}},
\oauthor{\bsnm{Fairbanks}, \binits{L.}},
\oauthor{\bsnm{Sanchez}, \binits{R.}},
\oauthor{\bsnm{Wacharanan}, \binits{P.}},
\oauthor{\bsnm{Abarbanel}, \binits{H.}}:
Data driven regional weather forecasting.
preprint,
Predictability, probabilistic forecasts, data assimilation, inverse
  problems/Climate, atmosphere, ocean, hydrology, cryosphere, biosphere/Big
  data and artificial intelligence
(November 2022).
\doiurl{10.5194/egusphere-2022-1222}
\end{botherref}
\endbibitem

\bibitem[\protect\citeauthoryear{Lorenz}{1963}]{lorenz1963deterministic}
\begin{barticle}
\bauthor{\bsnm{Lorenz}, \binits{E.N.}}:
\batitle{Deterministic nonperiodic flow}.
\bjtitle{Journal of atmospheric sciences}
\bvolume{20}(\bissue{2}),
\bfpage{130}--\blpage{141}
(\byear{1963})
\end{barticle}
\endbibitem

\bibitem[\protect\citeauthoryear{Sugihara and
  May}{1990}]{sugihara1990nonlinear}
\begin{barticle}
\bauthor{\bsnm{Sugihara}, \binits{G.}},
\bauthor{\bsnm{May}, \binits{R.M.}}:
\batitle{Nonlinear forecasting as a way of distinguishing chaos from
  measurement error in time series}.
\bjtitle{Nature}
\bvolume{344}(\bissue{6268}),
\bfpage{734}--\blpage{741}
(\byear{1990})
\end{barticle}
\endbibitem

\bibitem[\protect\citeauthoryear{Pedregosa et~al.}{2011}]{pedregosa2011scikit}
\begin{barticle}
\bauthor{\bsnm{Pedregosa}, \binits{F.}},
\bauthor{\bsnm{Varoquaux}, \binits{G.}},
\bauthor{\bsnm{Gramfort}, \binits{A.}},
\bauthor{\bsnm{Michel}, \binits{V.}},
\bauthor{\bsnm{Thirion}, \binits{B.}},
\bauthor{\bsnm{Grisel}, \binits{O.}},
\bauthor{\bsnm{Blondel}, \binits{M.}},
\bauthor{\bsnm{Prettenhofer}, \binits{P.}},
\bauthor{\bsnm{Weiss}, \binits{R.}},
\bauthor{\bsnm{Dubourg}, \binits{V.}}, \betal:
\batitle{Scikit-learn: Machine learning in python}.
\bjtitle{The Journal of Machine Learning Research}
\bvolume{12},
\bfpage{2825}--\blpage{2830}
(\byear{2011})
\end{barticle}
\endbibitem

\bibitem[\protect\citeauthoryear{Qin and Badgwell}{2003}]{qin_survey_2003}
\begin{barticle}
\bauthor{\bsnm{Qin}, \binits{S.J.}},
\bauthor{\bsnm{Badgwell}, \binits{T.A.}}:
\batitle{A survey of industrial model predictive control technology}.
\bjtitle{Control Engineering Practice}
\bvolume{11}(\bissue{7}),
\bfpage{733}--\blpage{764}
(\byear{2003})
\doiurl{10.1016/S0967-0661(02)00186-7}
\end{barticle}
\endbibitem

\bibitem[\protect\citeauthoryear{Fiedler et~al.}{2023}]{fiedler_-mpc_2023}
\begin{barticle}
\bauthor{\bsnm{Fiedler}, \binits{F.}},
\bauthor{\bsnm{Karg}, \binits{B.}},
\bauthor{\bsnm{Lüken}, \binits{L.}},
\bauthor{\bsnm{Brandner}, \binits{D.}},
\bauthor{\bsnm{Heinlein}, \binits{M.}},
\bauthor{\bsnm{Brabender}, \binits{F.}},
\bauthor{\bsnm{Lucia}, \binits{S.}}:
\batitle{do-mpc: {Towards} {FAIR} nonlinear and robust model predictive
  control}.
\bjtitle{Control Engineering Practice}
\bvolume{140},
\bfpage{105676}
(\byear{2023})
\doiurl{10.1016/j.conengprac.2023.105676}
\end{barticle}
\endbibitem

\bibitem[\protect\citeauthoryear{Andersson
  et~al.}{2019}]{andersson_casadi_2019}
\begin{barticle}
\bauthor{\bsnm{Andersson}, \binits{J.A.E.}},
\bauthor{\bsnm{Gillis}, \binits{J.}},
\bauthor{\bsnm{Horn}, \binits{G.}},
\bauthor{\bsnm{Rawlings}, \binits{J.B.}},
\bauthor{\bsnm{Diehl}, \binits{M.}}:
\batitle{{CasADi}: a software framework for nonlinear optimization and optimal
  control}.
\bjtitle{Mathematical Programming Computation}
\bvolume{11}(\bissue{1}),
\bfpage{1}--\blpage{36}
(\byear{2019})
\doiurl{10.1007/s12532-018-0139-4}
\end{barticle}
\endbibitem

\bibitem[\protect\citeauthoryear{Wächter and
  Biegler}{2006}]{wachter_implementation_2006}
\begin{barticle}
\bauthor{\bsnm{Wächter}, \binits{A.}},
\bauthor{\bsnm{Biegler}, \binits{L.T.}}:
\batitle{On the implementation of an interior-point filter line-search
  algorithm for large-scale nonlinear programming}.
\bjtitle{Mathematical Programming}
\bvolume{106}(\bissue{1}),
\bfpage{25}--\blpage{57}
(\byear{2006})
\doiurl{10.1007/s10107-004-0559-y}
\end{barticle}
\endbibitem

\bibitem[\protect\citeauthoryear{Mulansky and
  Kreuz}{2016}]{mulansky_pyspikepython_2016}
\begin{barticle}
\bauthor{\bsnm{Mulansky}, \binits{M.}},
\bauthor{\bsnm{Kreuz}, \binits{T.}}:
\batitle{{PySpike}—{A} {Python} library for analyzing spike train synchrony}.
\bjtitle{SoftwareX}
\bvolume{5},
\bfpage{183}--\blpage{189}
(\byear{2016})
\doiurl{10.1016/j.softx.2016.07.006}
\end{barticle}
\endbibitem

\bibitem[\protect\citeauthoryear{Steinmetz
  et~al.}{2021}]{steinmetz2021neuropixels}
\begin{barticle}
\bauthor{\bsnm{Steinmetz}, \binits{N.A.}},
\bauthor{\bsnm{Aydin}, \binits{C.}},
\bauthor{\bsnm{Lebedeva}, \binits{A.}},
\bauthor{\bsnm{Okun}, \binits{M.}},
\bauthor{\bsnm{Pachitariu}, \binits{M.}},
\bauthor{\bsnm{Bauza}, \binits{M.}},
\bauthor{\bsnm{Beau}, \binits{M.}},
\bauthor{\bsnm{Bhagat}, \binits{J.}},
\bauthor{\bsnm{B{\"o}hm}, \binits{C.}},
\bauthor{\bsnm{Broux}, \binits{M.}}, \betal:
\batitle{Neuropixels 2.0: A miniaturized high-density probe for stable,
  long-term brain recordings}.
\bjtitle{Science}
\bvolume{372}(\bissue{6539}),
\bfpage{4588}
(\byear{2021})
\end{barticle}
\endbibitem

\bibitem[\protect\citeauthoryear{Bottjer
  et~al.}{1986}]{bottjer_zebra_finch_hvc_1986}
\begin{barticle}
\bauthor{\bsnm{Bottjer}, \binits{S.W.}},
\bauthor{\bsnm{Miesner}, \binits{E.A.}},
\bauthor{\bsnm{Arnold}, \binits{A.P.}}:
\batitle{Changes in neuronal number, density and size account for increases in
  volume of song-control nuclei during song development in zebra finches}.
\bjtitle{Neuroscience Letters}
\bvolume{67}(\bissue{3}),
\bfpage{263}--\blpage{268}
(\byear{1986})
\doiurl{10.1016/0304-3940(86)90319-8}
\end{barticle}
\endbibitem

\bibitem[\protect\citeauthoryear{Milias-Argeitis and
  Khammash}{2015}]{milias-argeitis_adaptive_2015}
\begin{bchapter}
\bauthor{\bsnm{Milias-Argeitis}, \binits{A.}},
\bauthor{\bsnm{Khammash}, \binits{M.}}:
\bctitle{Adaptive model predictive control of an optogenetic system}.
In: \bbtitle{2015 54th {IEEE} {Conference} on {Decision} and {Control}
  ({CDC})},
pp. \bfpage{1265}--\blpage{1270}.
\bpublisher{IEEE},
\blocation{Osaka}
(\byear{2015}).
\doiurl{10.1109/CDC.2015.7402385}
\end{bchapter}
\endbibitem

\bibitem[\protect\citeauthoryear{Fox et~al.}{2023}]{fox_bayesian_2023}
\begin{barticle}
\bauthor{\bsnm{Fox}, \binits{Z.R.}},
\bauthor{\bsnm{Batt}, \binits{G.}},
\bauthor{\bsnm{Ruess}, \binits{J.}}:
\batitle{Bayesian filtering for model predictive control of stochastic gene
  expression in single cells}.
\bjtitle{Physical Biology}
\bvolume{20}(\bissue{5}),
\bfpage{055003}
(\byear{2023})
\doiurl{10.1088/1478-3975/ace094}
\end{barticle}
\endbibitem

\bibitem[\protect\citeauthoryear{Bemporad and
  Morari}{1999}]{bemporad2007robust}
\begin{bchapter}
\bauthor{\bsnm{Bemporad}, \binits{A.}},
\bauthor{\bsnm{Morari}, \binits{M.}}:
\bctitle{Robust model predictive control: A survey}.
In: \beditor{\bsnm{Garulli}, \binits{A.}},
\beditor{\bsnm{Tesi}, \binits{A.}} (eds.)
\bbtitle{Robustness in Identification and Control},
pp. \bfpage{207}--\blpage{226}.
\bpublisher{Springer},
\blocation{London}
(\byear{1999})
\end{bchapter}
\endbibitem

\bibitem[\protect\citeauthoryear{Smith et~al.}{2010}]{smith2010state}
\begin{barticle}
\bauthor{\bsnm{Smith}, \binits{A.C.}},
\bauthor{\bsnm{Scalon}, \binits{J.D.}},
\bauthor{\bsnm{Wirth}, \binits{S.}},
\bauthor{\bsnm{Yanike}, \binits{M.}},
\bauthor{\bsnm{Suzuki}, \binits{W.A.}},
\bauthor{\bsnm{Brown}, \binits{E.N.}}:
\batitle{State-space algorithms for estimating spike rate functions}.
\bjtitle{Computational Intelligence and Neuroscience}
\bvolume{2010},
\bfpage{1}--\blpage{14}
(\byear{2010})
\end{barticle}
\endbibitem

\bibitem[\protect\citeauthoryear{Fehrman and Meliza}{2024}]{fehrman2024model}
\begin{botherref}
\oauthor{\bsnm{Fehrman}, \binits{C.}},
\oauthor{\bsnm{Meliza}, \binits{C.D.}}:
Model predictive control of the neural manifold.
arXiv preprint arXiv:2406.14801
(2024)
\end{botherref}
\endbibitem

\bibitem[\protect\citeauthoryear{Langdon et~al.}{2023}]{langdon2023unifying}
\begin{botherref}
\oauthor{\bsnm{Langdon}, \binits{C.}},
\oauthor{\bsnm{Genkin}, \binits{M.}},
\oauthor{\bsnm{Engel}, \binits{T.A.}}:
A unifying perspective on neural manifolds and circuits for cognition.
Nature Reviews Neuroscience,
1--15
(2023)
\end{botherref}
\endbibitem

\bibitem[\protect\citeauthoryear{Lambeth et~al.}{2023}]{lambeth_robust_2023}
\begin{bchapter}
\bauthor{\bsnm{Lambeth}, \binits{K.}},
\bauthor{\bsnm{Singh}, \binits{M.}},
\bauthor{\bsnm{Sharma}, \binits{N.}}:
\bctitle{Robust control barrier functions for safety using a hybrid
  neuroprosthesis}.
In: \bbtitle{2023 {American} {Control} {Conference} ({ACC})},
pp. \bfpage{54}--\blpage{59}.
\bpublisher{IEEE},
\blocation{San Diego, CA, USA}
(\byear{2023}).
\doiurl{10.23919/ACC55779.2023.10155862}
\end{bchapter}
\endbibitem

\bibitem[\protect\citeauthoryear{Wolf and
  Schearer}{2022}]{wolf_trajectory_2022}
\begin{barticle}
\bauthor{\bsnm{Wolf}, \binits{D.N.}},
\bauthor{\bsnm{Schearer}, \binits{E.M.}}:
\batitle{Trajectory optimization and model predictive control for functional
  electrical stimulation-controlled reaching}.
\bjtitle{IEEE Robotics and Automation Letters}
\bvolume{7}(\bissue{2}),
\bfpage{3093}--\blpage{3098}
(\byear{2022})
\doiurl{10.1109/LRA.2022.3145946}
\end{barticle}
\endbibitem

\bibitem[\protect\citeauthoryear{Singh and
  Sharma}{2023}]{singh_data-driven_2023}
\begin{bchapter}
\bauthor{\bsnm{Singh}, \binits{M.}},
\bauthor{\bsnm{Sharma}, \binits{N.}}:
\bctitle{Data-driven model predictive control for drop foot correction}.
In: \bbtitle{2023 {American} {Control} {Conference} ({ACC})},
pp. \bfpage{2615}--\blpage{2620}.
\bpublisher{IEEE},
\blocation{San Diego, CA, USA}
(\byear{2023}).
\doiurl{10.23919/ACC55779.2023.10156600}
\end{bchapter}
\endbibitem

\bibitem[\protect\citeauthoryear{Bao et~al.}{2019}]{bao_model_2019}
\begin{barticle}
\bauthor{\bsnm{Bao}, \binits{X.}},
\bauthor{\bsnm{Kirsch}, \binits{N.}},
\bauthor{\bsnm{Dodson}, \binits{A.}},
\bauthor{\bsnm{Sharma}, \binits{N.}}:
\batitle{Model predictive control of a feedback-linearized hybrid
  neuroprosthetic system with a barrier penalty}.
\bjtitle{Journal of Computational and Nonlinear Dynamics}
\bvolume{14}(\bissue{10}),
\bfpage{101009}
(\byear{2019})
\doiurl{10.1115/1.4042903}
\end{barticle}
\endbibitem

\bibitem[\protect\citeauthoryear{Chatterjee
  et~al.}{2020}]{chatterjee2020fractional}
\begin{barticle}
\bauthor{\bsnm{Chatterjee}, \binits{S.}},
\bauthor{\bsnm{Romero}, \binits{O.}},
\bauthor{\bsnm{Ashourvan}, \binits{A.}},
\bauthor{\bsnm{Pequito}, \binits{S.}}:
\batitle{Fractional-order model predictive control as a framework for
  electrical neurostimulation in epilepsy}.
\bjtitle{Journal of Neural Engineering}
\bvolume{17}(\bissue{6}),
\bfpage{066017}
(\byear{2020})
\end{barticle}
\endbibitem

\bibitem[\protect\citeauthoryear{Brar et~al.}{2018}]{brar2018seizure}
\begin{bchapter}
\bauthor{\bsnm{Brar}, \binits{H.K.}},
\bauthor{\bsnm{Exarchos}, \binits{I.}},
\bauthor{\bsnm{Pan}, \binits{Y.}},
\bauthor{\bsnm{Theodorou}, \binits{E.}},
\bauthor{\bsnm{Mahmoudi}, \binits{B.}}:
\bctitle{Seizure reduction using model predictive control}.
In: \bbtitle{2018 40th Annual International Conference of the IEEE Engineering
  in Medicine and Biology Society (EMBC)},
pp. \bfpage{3152}--\blpage{3155}
(\byear{2018}).
\bcomment{IEEE}
\end{bchapter}
\endbibitem

\end{thebibliography}
\end{document}